\documentstyle[12pt]{article}
\begin{document}
\begin{center}
{\large {\bf FINITE-DIMENSIONAL REPRESENTATIONS \\
OF THE QUANTUM SUPERALGEBRA U$_{q}$[gl(2/2)]:\\
  I. Typical representations at generic $q$}} \\
\vskip 1truecm
\def\thefootnote{*)}
{\bf Nguyen Anh Ky}\hspace*{1.5mm}\footnote{Permanent Mailing Address:
Centre for Polytechnic Educations (Trung t\^{a}m
gi\'{a}o d\d{u}c k\~{y} thu\d{\^{a}}t t\^{o}ng h\d{o}p), Vinh, {\bf
Vietnam}}
\def\thefootnote{ }
\footnote{Address after June 01, 1993: Institute for Nuclear
Research and Nuclear Energy, Boul. Tsarigradsko chaussee 72, Sofia 1784,
{\bf Bulgaria}}\\[1mm]

\normalsize International Centre for Theoretical Physics\\P.O. Box 586,
Trieste 34100, {\bf Italy}
\vskip 2truecm
{\bf Abstract} \\[1mm]
\end{center}

   In the present paper we construct all typical finite-dimensional
representations of the quantum Lie superalgebra $U_{q}[gl(2/2)]$ at
generic deformation parameter $q$. As in the non-deformed case the
finite-dimensional $U_{q}[gl(2/2)]$-module $W^{q}$ obtained
is irreducible and can be decomposed into finite-dimensional
irreducible $U_{q}[gl(2)\oplus gl(2)]$-submodules $V^{q}_{k}$.\\[1.2cm]
PACS numbers: ~ 02.20Tw, 11.30Pb.
\newpage
\begin{flushleft}
{\large {\bf 1. Introduction}}
\end{flushleft}
\vspace*{1mm}

  Since the quantum deformations $^{1-5}$ became a subject of
intensive investigations many (algebraic and geometric)
structures and different representations of quantum (super-)
groups have been obtained and understood. For instance, the
quantum algebra $U_{q}[sl(2)]$ is very well studied $^{6-8}$.
Originated from intensive investigations on the quantum inverse
scattering method and the Yang-Baxter equations, the quantum
groups have found various applications in theoretical physics and
mathematics (see in this context, for example, Refs. $^{9-12}$).
As in the non-deformed case for applications of quantum groups we
often need their explicit representations. Being a subject of
many investigations, representations of quantum groups,
especially representations of quantum superalgebras are presently
under development. However, although the progress in this
direction is remarkable the problem is still far from being
satisfactorily solved. Explicit representations are known only
for  quantum Lie superalgebras of lower ranks and of particular
types like $U_{q}[osp(1/2)]$ $^{13}$, $U_{q}[gl(1/n)]$ $^{14}$,
etc.. For higher rank quantum Lie superalgebras $^{15-18}$,
besides some q-oscillator representations which are most popular
among those constructed, we do not know so much about other
representations, in particular the finite-dimensional ones. Some
general aspects and module structures of finite-dimensional
representations of $U_{q}[gl(m/n)]$ are considered in Ref.
$^{18}$ (see also Ref. $^{14}$) but without their explicit
constructions.  So the question concerning an explicit
construction of finite-dimensional representations of
$U_{q}[gl(m/n)]$ is still unsolved for $m$ and $n$ $\geq$ 2.\\

  Here, extending the method developed by Kac $^{19}$ in the case
of Lie superalgebras (from now on, only superalgebras) we shall
construct all finite-dimensional representations of the quantum
Lie superalgebra $U_{q}[gl(2/2)]$ at generic $q$, i.e. $q$ is not
a root of unity. It turns out that the finite-dimensional
$U_{q}[gl(2/2)]$-modules have similar structures to that of the
non-deformed ones $^{20,21}$ and are decomposed into
finite-dimensional irreducible $U_{q}[gl(2)\oplus
gl(2)]$-modules. Finite-dimensional $U_{q}[gl(2/2)]$-modules can
be classified again as  typical or atypical ones (see the {\it
Proposition} 2). In the frame-work of this paper for the sake of
simplicity we shall consider only the typical representations at
generic $q$. When $q$ is a root of unity, as emphasized also in
$^{18}$, the structures of $U_{q}[gl(2/2)]$-modules are
drastically different in comparison with the structures of
$gl(2/2)$-modules $^{20,21}$. The present investigation on typical
representations at generic $q$ is easily extended on atypical
representations at generic $q$ $^{22}$ and finite-dimensional
representations at $q$ being a root of unity $^{23}$.\\

  The paper is organized as follows. In order to make the present
construction clear, in the section 2 we expose some introductory
concepts and basic definitions of quantum superalgebras,
especially $U_{q}[gl(m/n)]$.  We also describe briefly the
procedure used for constructing finite-dimensional
representations of $U_{q}[gl(m/n)]$. The quantum superalgebra
$U_{q}[gl(2/2)]$ is defined in section 3. The section 4 is
devoted to construction of finite-dimensional representations of
$U_{q}[gl(2/2)]$. Some comments and the conclusion are made in
the section 5, while the references are given in the last section
6.\\

 For a convenient reading we shall keep as many as possible the
abbreviations and notations used in Ref. $^{20}$ among the
following ones:
\begin{tabbing}
\=12345678\=$V_{l}\otimes V_{r}$ - \=tensor product between two linear spaces
$V_{l}$ and $V_{r}$\= or a tensor product\=\kill
\>\>fidirmod(s) - finite-dimensional irreducible module(s),\\[2mm]
\>\>GZ basis - Gel'fand-Zetlin basis,\\[2mm]
\>\>lin.env.\{X\} - linear envelope of X,\\[2mm]
\>\>$q$ - the deformation parameter,\\[2mm]
\>\>$V^{q}_{l}\otimes V^{q}_{r}$ - tensor product between two linear spaces
$V^{q}_{l}$ and $V^{q}_{r}$\\
\>\>\>~~or a tensor product between a $U_{q}[gl(2)_{l}]$-module $V^{q}_{l}$\\
\>\>\>~~and a $U_{q}[gl(2)_{r}]$-module $V^{q}_{r}$,\\[2mm]
\>\>$T^{q}\odot V^{q}_{0}$ - tensor product between two $U_{q}[gl(2)\oplus
gl(2)]$-modules\\
\>\>\>~~$T^{q}$ and $V^{q}_{0}$,\\[2mm]
\>\>$[x]_{q}={q^{x}-q^{-x}\over q-q^{-1}}$,~ where $x$ is some
number or operator,\\[2mm]
\>\>$[x]\equiv [x]_{q^{2}}$,\\[2mm]
\>\>$[E,F\}$ - supercommutator between $E$ and $F$,\\[2mm]
\>\>$[E,F]_{q}\equiv EF-qFE$ - q-deformed commutator between $E$
and $F$,\\[2mm]
\>\>$a_{ij}$ - an element of the Cartan matrix $(a_{ij})$,\\[2mm]
\>\>$q_{i}=q^{d_{i}}$,~ where $d_{i}$ are rational numbers such that\\
\>\>$d_{i}a_{ij} = d_{j}a_{ji}$, $i,j=1,2,...,r$,\\
\>\>${\cal E}_{i}=e_{i}q_{i}^{-h_{i}}\equiv e_{i}k_{i}^{-1}k_{i+1}$,\\
\>\>${\cal F}_{i}=f_{i}q_{i}^{-h_{i}}\equiv f_{i}k_{i}^{-1}k_{i+1}$.
\end{tabbing} Note that we must not confuse the quantum
deformation $[x]\equiv [x]_{q^{2}}$ of $x$ with the highest
weight (signature) $[m]$ in the GZ basis $(m)$ or with the
notation [ , ] for commutators.\\[7mm]
{\large {\bf 2. Some
introductory concepts of quantum superalgebras}}\\

   Let $g$ be a rank $r$ (semi-) simple superalgebra, for
example, $sl(m/n)$ or $osp(m/n)$. The quantum superalgebra
$U_{q}(g)$ as a quantum deformation (q-deformation) of the
universal enveloping algebra $U(g)$ of $g$, is completely defined
by the Cartan-Chevalley canonical generators $h_{i}$, $e_{i}$ and
$f_{i}$, $i=1,2,...,r$ which satisfy $^{15-17}$\\[1mm]

 a) the quantum Cartan-Kac supercommutation relations
\begin{tabbing}
\=11111111111111111111111111111\=$[h_{i},h_{j}]$\= =
0\=2222222222222222222222222222222\= \kill
\>\>$[h_{i},h_{j}]$\> = 0,\\[1mm]
\>\>$[h_{i},e_{j}]$\> = $a_{ij}e_{j}$,\\[1mm]
\>\>$[h_{i},f_{j}]$\> = $-a_{ij}f_{j}$,\\[1mm]
\>\>$[e_{i},f_{j}\}$\> = $\delta_{ij}[h_{i}]_{q_{i}^{2}}$,\>\>(2.1)
\end{tabbing}

 b) the quantum Serre relations $$(ad_{q}{\cal
E}_{i})^{1-\tilde{a}_{ij}}{\cal E}_{j}=0,$$ $$(ad_{q}{\cal
F}_{i})^{1-\tilde{a}_{ij}}{\cal F}_{j}=0\eqno(2.2)$$ where
$(\tilde{a}_{ij})$ is a matrix obtained from the non-symmetric
Cartan matrix $(a_{ij})$ by replacing the strictly positive
elements in the rows with 0 on the diagonal entry by $-1$, while
$ad_{q}$ is the q-deformed adjoint operator given by the formula
(2.8)\\[2mm] and\\[2mm]

 c) the quantum extra-Serre relations $^{24-26 }$ (for $g$ being
$sl(m/n)$ or $osp(m/n)$)
$$\{[e_{m-1},e_{m}]_{q^{2}},[e_{m},e_{m+1}]_{q^{2}}\}=0,$$
$$\{[f_{m-1},f_{m}]_{q^{2}},[f_{m},f_{m+1}]_{q^{2}}\}=0,\eqno(2.3)$$
being additional constraints on the unique odd Chevalley
generators $e_{m}$ and $f_{m}$. In the above formulas we denoted
$q_{i} = q^{d_{i}}$ where $d_{i}$ are rational numbers
symmetrizing the Cartan matrix $d_{i}a_{ij}=d_{j}a_{ji}$, $1\leq
i,j\leq r$. For example, in case $g=sl(m/n)$ we have
$$d_{i}=\left\{
\begin{array}{ll}
1 & ~~~~ if ~~ 1\leq i\leq m,\\ -1 & ~~~~ if ~~ m+1\leq i\leq
r=m+n-1.
\end{array}\right.
\eqno(2.4)$$

  The above-defined  quantum superalgebras form a subclass of a
special class of Hopf algebras called by Drinfel'd
quasitriangular Hopf algebras $^{2}$.  They are endowed with a
Hopf algebra structure given by the following additional maps:\\

a) {\it coproduct} $\Delta$ : ~~$U$ $\rightarrow$ $U\otimes U$
\begin{tabbing}
\=111111111111111111111111\=$\Delta(h_{i})$\= $= h_{i}\otimes 1 +
1\otimes h_{i}$\=22222222222222222\=2222222\= \kill
\>\>$\Delta(1)$\> $= 1\otimes 1$,\>\>\>\\[2mm]
\>\>$\Delta(h_{i})$\> $= h_{i}\otimes 1 + 1\otimes h_{i}$,\>\>\>\\[2mm]
\>\>$\Delta(e_{i})$\> $= e_{i}\otimes q_{i}^{h_{i}} +
q_{i}^{-h_{i}}\otimes e_{i}$,\>\>\>\\[2mm]
\>\>$\Delta(f_{i})$\> $= f_{i}\otimes q_{i}^{h_{i}} +
q_{i}^{-h_{i}}\otimes f_{i}$,\>\>\>(2.5)
\end{tabbing}

b) {\it antipode} $S$ : ~~~$U$ $\rightarrow$ $U$
\begin{tabbing}
\=111111111111111111111111\=$S(h_{i})$\=
$= -h_{i}$\=22222222222222222\=22222222222222222\= \kill
\>\>$S(1)$\> $= 1$,\>\>\>\\[2mm]
\>\>$S(h_{i})$\> $= -h_{i}$,\>\>\>\\[2mm]
\>\>$S(e_{i})$\> $= -q^{a_{ii}}_{i}e_{i}$,\>\>\>\\[2mm]
\>\>$S(f_{i})$\> $= -q^{-a_{ii}}_{i}f_{i}$ \>\>\>(2.6)
\end{tabbing}
\vspace*{2mm}
and\\

c) {\it counit} $\varepsilon$ : ~$U$ $\rightarrow$ $C$
\begin{tabbing}
\=111111111111111111111111\=$S(h_{i})$\=
$= -h_{i}$2222222222222222222222222222222222\= \kill
\>\>$\varepsilon(1)$\> $= 1$,\>\\[2mm]
\>\>$\varepsilon(h_{i})$\>
$=\varepsilon(e_{i})=\varepsilon(f_{i})=0$,\>(2.7)
\end{tabbing}
Then the quantum adjoint operator $ad_{q}$ has the following form
$^{16,27}$
$$ad_{q} = (\mu_{L} \otimes \mu_{R})(id \otimes
S)\Delta\eqno(2.8)$$ with  $\mu_{L}$ (respectively, $\mu_{R}$)
being the left (respectively, right) multiplication: $\mu_{L}(x)y
= xy$ (respectively, $\mu_{R}(x)y = (-1)^{degx.degy} yx$).\\

  A quantum superalgebra $U_{q}[gl(m/n)]$ is generated by the
generators $k_{i}^{\pm 1}\equiv q_{i}^{\pm E_{ii}}$, $e_{j}\equiv
E_{j,j+1}$ and $f_{j}\equiv E_{j+1,j}$, $i=1,2,...,m+n$,
$j=1,2,...,m+n-1$ such that the following relations hold (cf.
Refs. $^{14,18}$)\\

  a) the super-commutation relations
\begin{tabbing}
\=12345678912345
\=$k_{i}e_{j}k_{i}^{-1}$1\==1\=$q_{i}^{(\delta_{ij}-\delta_{i,j+1})}e_{j}$,~~~~
\=$k_{i}f_{j}k_{i}^{-1}$1\==12\=$q_{i}^{(\delta_{ij+1}-\delta_{i,j})}f_{j}$,
1234567\=\kill
\>\>$k_{i}k_{j}$\>=\>$k_{j}k_{i}$~,\>$k_{i}k_{i}^{-1}$\>=\>$k_{i}^{-1}k_{i}$ =
1~,\\[1mm]
\>\>$k_{i}e_{j}k_{i}^{-1}$\>=\>$q_{i}^{(\delta_{ij}-\delta_{i,j+1})}e_{j}$~,
\>$k_{i}f_{j}k_{i}^{-1}$\>=\>$q_{i}^{(\delta_{ij+1}-\delta_{i,j})}f_{j}$~,
\\[1mm]
\>\>$[e_{i},f_{j}\}$\>=\>$\delta_{ij}[h_{i}]_{q^{2}_{i}}$, ~where
\>~~$q_{i}^{h_{i}}$\>=\>$k_{i}k_{i+1}^{-1}$,\>(2.9)
\end{tabbing}

  b) the Serre relations (2.2) taking now the explicit forms
\begin{tabbing}
\=1234567891234.\=$[e_{i},e_{j}]$\=~=~$[f_{i},f_{j}]$\=~=~0,~~if
$|i-j|\neq 1$123\=23456789123456789123~.\=\kill
\>\>$[e_{i},e_{j}]$\>~=~$[f_{i},f_{j}]$\>~=~0,~~if $|i-j|\neq 1$,\\[1mm]
\>\>~~$e_{m}^{2}$\>~=~~~$f_{m}^{2}$\>~=~0,\\[1mm]
\>\>$[e_{i},[e_{i},e_{j}]_{q^{\pm 2}}]_{q^{\mp 2}}$\>
\>~=~~$[f_{i},[f_{i},f_{j}]_{q^{\pm 2}}]_{q^{\mp 2}}$~=\>~0, ~~if
$|i-j|=1$\>(2.10)\\[1mm] \end{tabbing}
\vspace*{2mm}
and\\

  c) the extra-Serre relations (2.3)
$$\{[e_{m-1},e_{m}]_{q^{2}},[e_{m},e_{m+1}]_{q^{2}}\}~= 0,$$
$$\{[f_{m-1},f_{m}]_{q^{2}},[f_{m},f_{m+1}]_{q^{2}}\}~=
0.\eqno(2.11)$$ Here, besides $d_{i}$, $1\leq i\leq r=m+n-1$ given
in (2.4) we introduced $d_{m+n}=-1$.
The Hopf structure on $k_{i}$ looks as
\begin{tabbing}
\=1234567891234567891234567891\=$\Delta(k_{i})$~\= =~ \=$k_{i}\otimes
k_{i}$1234567891234567891234567\=\kill
\>\>$\Delta(k_{i})$\> = \>$k_{i}\otimes k_{i}$,\\[1mm]
\>\>$S(k_{i})$\> = \>$k_{i}^{-1}$,\\[1mm]
\>\>$\varepsilon(k_{i})$\> = \>1.\>(2.12)
\end{tabbing}

  The generators $E_{ii}$, $E_{i,i+1}$ and $E_{i+1,i}$ together
with the generators defined in the following way
\begin{tabbing}
\=123456789\=$E_{i,j+1}$\=:= $[E_{ik}E_{kj}]_{q^{-2}}$x\= $\equiv
E_{ik}E_{kj}~-~q^{-2}E_{kj}E_{ik}$,
\=$~~i<k<j$,\=12345676~\=\kill
\>\>$E_{ij}$\>:= ~$[E_{ik}E_{kj}]_{q^{-2}}$\> $\equiv
{}~E_{ik}E_{kj}~-~q^{-2}E_{kj}E_{ik}$,\>$~~i<k<j$,
\\[1mm]
\>\>$E_{ji}$\>:= ~$[E_{jk}E_{ki}]_{q^{2}}$\> $\equiv
{}~E_{jk}E_{ki}~-~q^{2}E_{ki}E_{jk}$$,~$\>$~~i<k<j$,\>\> (2.13)
\end{tabbing}
play an analogous role as the Weyl generators $e_{ij}$,
$$(e_{ij})_{kl} = \delta_{ik}\delta_{jl}, \eqno(2.14)$$ of the
superalgebra $gl(m/n)$ whose universal enveloping algebra
$U[gl(m/n)]$ represents a classical limit of $U_{q}[gl(m/n)]$
when $q\rightarrow 1$.\\

    The quantum algebra $U_{q}[gl(m/n)_{0}]\cong
U_{q}[gl(m)\oplus gl(n)]$ generated by $k_{i}$, $e_{j}$ and
$f_{j}$, $i=1,2,...,m+n$, $m\neq j=1,2,...,m+n-1$,
$$U_{q}[gl(m/n)_{0}] ~=~lin.env.\{E_{ij}\|~ 1\leq i,j\leq m
{}~~and~~ m+1\leq i,j\leq m+n\}\eqno(2.15)$$ is an even subalgebra
of $U_{q}[gl(m/n)]$. Note that $U_{q}[gl(m/n)_{0}]$ is included
in the largest even subalgebra $U_{q}[gl(m/n)]_{0}$ containing
all elements of $U_{q}[gl(m/n)]$ with even powers of the odd
generators.\\

  As is shown by M. Rosso $^{28}$ and C. Lusztig $^{29}$, a
finite-dimensional representation of a Lie algebra $g$ can be
deformed to a finite-dimensional representation of its quantum
analogue $U_{q}(g)$. In particular, finite-dimensional
representations  of $U_{q}[gl(m)\oplus gl(n)]$ are quantum
deformations of those of $gl(m)\oplus gl(n)$. Hence, a
finite-dimensional irreducible representation of
$U_{q}[gl(m)\oplus gl(n)]$ is again highest weight. Following the
classical procedure $^{19,20}$ we can construct representations
of $U_{q}[gl(m/n)]$ induced from finite-dimensional irreducible
representations of $U_{q}[gl(m)\oplus gl(n)]$ which, as we can
see from (2.9-11) and (2.15), is  the stability subalgebra of
$U_{q}[gl(m/n)]$. Let $V_{0}^{q}(\Lambda)$ be a
$U_{q}[gl(m)\oplus gl(n)]$-fidirmod characterized by some highest
weight $\Lambda$. For a basis of $V_{0}^{q}(\Lambda)$ we can
choose the Gel'fand-Zetlin (GZ) tableaux $^{30}$, since the
latter are invariant under the quantum deformations
$^{28,29,31,32}$.  Therefore, the highest weight $\Lambda$ is
described again by the first row of the GZ tableaux called from
now on as the GZ (basis) vectors.\\

  Demanding $$E_{m,m+1}V_{0}^{q}(\Lambda)\equiv
e_{m}V_{0}^{q}(\Lambda)=0\eqno(2.16)$$ i.e.
$$U_{q}(A_{+})V_{0}^{q}(\Lambda)=0 \eqno(2.17)$$ we turn
$V_{0}^{q}(\Lambda)$ into a $U_{q}(B)$-module, where $$A_{+} =
\{E_{ij}\|~ 1\leq i\leq m < j\leq m+n\} \eqno(2.18)$$ $$B =
A_{+}\oplus gl(m)\oplus gl(n) \eqno(2.19)$$ The $U_{q}[gl(m/n)]$
-module $W^{q}$ induced from the $U_{q}[gl(m)\oplus
gl(n)]$-module $V_{0}^{q}(\Lambda)$ is the factor-space
$$W^{q}=W^{q}(\Lambda) = [U_{q} \otimes
V_{0}^{q}(\Lambda)]/I^{q}(\Lambda)
\eqno(2.20)$$
where $U_{q}\equiv U_{q}[gl(m/n)]$, while $I^{q}(\Lambda)$ is the
subspace $$I^{q}(\Lambda) = lin.env.\{ub\otimes v -u\otimes bv \|
{}~ u\in U_{q},~ b\in U_{q}(B)\subset U_{q},~ v\in
V_{0}^{q}(\Lambda)\} \eqno(2.21)$$

    In order to complete the present section let us  note that
the modules $W^{q}(\Lambda)$ and $V_{0}^{q}(\Lambda)$ have one
and the same highest vector. Therefore, they are characterized by
one and the same highest weight $\Lambda$.\\[7mm]
{\large {\bf 3.$U[gl(2/2)]$$U[gl(2/2)]$The quantum superalgebra
$U_{q}[gl(2/2)]$}}\\

  The quantum superalgebra $U_{q}\equiv U_{q}[gl(2/2)]$ is
generated by the generators $E_{ii}$, $i=1,2,3,4$, $E_{12}\equiv
e_{1}$, $E_{23}\equiv e_{2}$, $E_{34}\equiv e_{3}$, $E_{21}\equiv
f_{1}$, $E_{32}\equiv f_{2}$ and $E_{43}\equiv f_{3}$ satisfying
the relations (2.9-11) which now read
\begin{tabbing}
\=1234567891234567891\=$[E_{ii},E_{jj}]$12345\= =1\= 0,1234
\=$[E_{ii},E_{j,j+1}]$\= =
\=$(\delta_{ij}-\delta_{i,j+1})E_{j,j+1}$1\=\kill

{}~~~~a) the super-commutation relations ($1\leq i,i+1,j,j+1\leq
4$):\\[2mm]
\>\>$[E_{ii},E_{jj}]$\> = \>0,\\[1mm]
\>\>$[E_{ii},E_{j,j+1}]$\>=\>$(\delta_{ij}-\delta_{i,j+1})E_{j,j+1}$,\\[1mm]
\>\>$[E_{ii},E_{j+1,j}]$\>=\>$(\delta_{i,j+1}-\delta_{ij})E_{j+1,j}$,\\[1mm]
\>\>$[E_{i,i+1},E_{j+1,j}\}$\>=\>$\delta_{ij}[h_{i}]_{q^{2}}$,
{}~~$h_{i}=(E_{ii}-{d_{i+1}\over
d_{i}}E_{i+1,i+1}),$\>\>\>\>(3.1)\\[4mm] with
$d_{1}=d_{2}=-d_{3}=-d_{4}=1$,\\[4mm]

{}~~~~b) the Serre-relations:\\[2mm]
\>\>~~$[E_{12},E_{34}]$\>=\>$[E_{21},E_{43}]$\>~~~~~~~=~0,\\[1mm]
\>\>~~~~~~$E_{23}^{2}$\>=\>~~~~$E_{32}^{2}$\>~~~~~~~=~0,\\[1mm]
\>\>~~$[E_{12},E_{13}]_{q^{2}}$\>=\>$[E_{24},E_{34}]_{q^{2}}$\>~~~~~~~=~0,

\\[1mm]
\>\>~~$[E_{21},E_{31}]_{q^{2}}$\>=\>$[E_{42},E_{43}]_{q^{2}}$\>~~~~~~~=~0,
                                        \>\>\>(3.2)\\[2mm]
and\\[2mm]

{}~~~~c) the extra-Serre relations:\\[2mm]
\>\>~~~~$\{E_{13},E_{24}\}$\>=\>0,\\[1mm]
\>\>~~~~$\{E_{31},E_{42}\}$\>=\>0,\>\>\>\>(3.3)\\[2mm]
respectively. Here, for a further convenience, the operators
\\[2mm]
\>\>~~~~~~~~~~~~$E_{13}$~:\>=\>$[E_{12},E_{23}]_{q^{-2}}$,\\[1mm]
\>\>~~~~~~~~~~~~$E_{24}$~:\>=\>$[E_{23},E_{34}]_{q^{-2}}$,\\[1mm]
\>\>~~~~~~~~~~~~$E_{31}$~:\>=\>$-[E_{21},E_{32}]_{q^{-2}},$\\[1mm]
\>\>~~~~~~~~~~~~$E_{42}$~:\>=\>$-[E_{32},E_{43}]_{q^{-2}}$.\>\>\>\>(3.4)
\end{tabbing}
and the operators composed in the following way
\begin{tabbing}
\=123456789123456789\=$E_{41}$~\=:=1
\=$[E_{21},[E_{32},E_{43}]_{q^{-2}}]_{q^{-2}}$
\=$\equiv [E_{21},E_{42}]_{q^{-1}}$123456789123\=\kill
\>\>$E_{14}$~\>:=\>$[E_{12},[E_{23},E_{34}]_{q^{-2}}]_{q^{-2}}$
\>$\equiv ~[E_{12},E_{24}]_{q^{-2}}$,\\[1mm]
\>\>$E_{41}$\>:=\>$[E_{21},[E_{32},E_{43}]_{q^{-2}}]_{q^{-2}}$
\>$\equiv ~-[E_{21},E_{42}]_{q^{-2}}$\>(3.5)
\end{tabbing}
are defined as new generators. The latter are odd and have
vanishing squares. They, together with the Cartan-Chevalley
generators, form a full system of q-analogues of the Weyl
generators $e_{ij}$, $1\leq i,j\leq 4$, of  the superalgebra
$gl(2/2)$ whose universal enveloping algebra $U[gl(2/2)]$ is a
classical limit of $U_{q}[gl(2/2)]$ when $q\rightarrow 1$. Other
commutation relations between $E_{ij}$ follow from the relations
(3.1-3) and the definitions (3.4-5).\\

 The subalgebra $U_{q}[gl(2/2)_{0}]~ (\subset
U_{q}[gl(2/2)]_{0}\subset U_{q}[gl(2/2)])$ is even and isomorphic
to $U_{q}[gl(2)\oplus gl(2)]\equiv U_{q}[gl(2)]\oplus
U_{q}[gl(2)]$ which is completely defined by $E_{ii}$, $1\leq
i\leq 4$, $E_{12}$, $E_{34}$, $E_{21}$ and $E_{43}$
$$U_{q}[gl(2/2)_{0}]~=~ lin.env.\{E_{ij}\|~
i,j=1,2~~and~~i,j=3,4\}\eqno(3.6)$$ In order to distinguish two
components of $U_{q}[gl(2/2)_{0}]$ we set
$$left~~U_{q}[gl(2)]\equiv U_{q}[gl(2)_{l}]:=lin.env.\{E_{ij}\|~
i,j=1,2\},\eqno(3.7)$$ $$right~U_{q}[gl(2)]\equiv
U_{q}[gl(2)_{r}]:=lin.env.\{E_{ij}\|~ i,j=3,4\}.\eqno(3.8)$$ That
means $$U_{q}[gl(2/2)_{0}]~=~U_{q}[gl(2)_{l}\oplus
gl(2)_{r}].\eqno(3.9)$$ Let $V^{q}(\Lambda)$ be a
$U_{q}[gl(2/2)_{0}]$-fidirmod of the highest weight $\Lambda$.
Thus  $V^{q}$ can be decomposed into a tensor product
$$V^{q}(\Lambda)=V_{l}^{q}(\Lambda_{l})\otimes
V_{r}^{q}(\Lambda_{r}),\eqno(3.10)$$ between a
$U_{q}[gl(2)_{l}]$-fidirmod $V_{l}^{q}(\Lambda_{l})$ of a highest
weight $\Lambda_{l}$ and a $U_{q}[gl(2)_{r}]$-fidirmod
$V_{r}^{q}(\Lambda_{r})$ of a highest weight $\Lambda_{r}$, where
$\Lambda_{l}$ and $\Lambda_{r}$ are defined respectively as the
left and right components of $\Lambda$ :
$$\Lambda=[\Lambda_{l},\Lambda_{r}].\eqno(3.11)$$\\[5mm] {\large
{\bf 4. Finite-dimensional representations of
$U_{q}[gl(2/2)]$}}\\

  Here, we shall construct finite-dimensional representations of
$U_{q}[gl(2/2)]$ induced from  finite-dimensional irreducible
representations of $U_{q}[gl(2/2)_{0}]$. In the frame-work of the
present paper we consider only typical representations of
$U_{q}[gl(2/2)]$ at generic $q$.  Atypical representations at
generic $q$ and finite-dimensional representations of
$U_{q}[gl(2/2)]$ at roots of unity are subjects of later
publications $^{22,23}$.\\

As mentioned earlier a fidirmod $V_{0}^{q}(\Lambda)$ of the
quantum algebra $U_{q}[gl(2/2)_{0}]$ represents a quantum
deformation (q-deformation) of some fidirmod $V_{0}(\Lambda)$ of
the algebra $gl(2/2)_{0}$. Moreover, following the classical
procedure we can construct $U_{q}[gl(2/2)]$-fidirmods induced
from $U_{q}[gl(2/2)_{0}]$-fidirmods.  Setting
$$E_{23}V_{0}^{q}=0,\eqno(4.1)$$ a $U_{q}[gl(2/2)]$-module
$W^{q}$ induced from the $U_{q}[gl(2/2)_{0}]$-fidirmod
$V_{0}^{q}$, by the construction, is the factor-space (2.20) with
$m=n=2$ : $$W^{q}(\Lambda) = [U_{q} \otimes
V_{0}^{q}(\Lambda)]/I^{q}(\Lambda),
\eqno(4.2)$$
where $$I^{q}(\Lambda) = lin.env.\{ub\otimes v -u\otimes bv \|
{}~u\in U_{q}, ~b\in U_{q}(B)\subset U_{q}, ~v\in
V_{0}^{q}(\Lambda)\}$$ $$U_{q}(B)=lin.env.\{E_{ij}, ~E_{23}\|
{}~i,j=1,2 ~~and~~ i,j=3,4\}\eqno(4.3)$$
Any vector $w$ from the
module $W^{q}$ has the form $$w=u\otimes v,~~~ u\in U_{q},~~ v\in
V_{0}^{q}\eqno(4.4)$$
Then $W^{q}$ is a $U_{q}[gl(2/2)]$-module
in the sense $$gw\equiv g(u\otimes v)=gu\otimes v\in
W^{q}\eqno(4.5)$$ for  $g$, $u\in U_{q}$, $w\in W^{q}$ and $v\in
V_{0}^{q}$.\\

  In the next two subsections we shall  construct the bases of
the module $W^{q}$ and find the explicit matrix elements for the
typical representations of $U_{q}[gl(2/2)]$.\\

  {\bf 4.1 The bases}\\

  Since the GZ basis is invariant under the q-deformation, for a
basis of a $U_{q}[gl(2)]$-fidirmod $V_{0}$ we can choose

$$
\left[
\begin{array}{c}
                        m_{12}~~~m_{22}\\ m_{11}
\end{array}
\right]
\equiv
\left[
\begin{array}{c}
                             [m]\\ m_{11}
\end{array}
\right]
\eqno(4.6)$$
where $m_{ij}$ are complex numbers such that $m_{12}-m_{11}\in
{\bf Z}_{+}$ and $m_{11}-m_{22}\in {\bf Z}_{+}$. Under the actions of the
$U_{q}[gl(2)]$-generators $E_{ij}$, $i,j=1,2$ the basis (4.6)
transforms as follows $^{32}$
\begin{eqnarray*}
{}~~~~~E_{11}\left[
\begin{array}{c}
                        m_{12}~~~m_{22}\\ m_{11}
\end{array}
\right]
& = &(l_{11}+1)\left[
\begin{array}{c}
                        m_{12}~~~m_{22}\\ m_{11}
\end{array}
\right],\\[2mm]
E_{22}\left[
\begin{array}{c}
                        m_{12}~~~m_{22}\\ m_{11}
\end{array}
\right]
& = &(l_{12}+l_{22}-l_{11}+2)\left[
\begin{array}{c}
                        m_{12}~~~m_{22}\\ m_{11}
\end{array}
\right],\\[2mm]
E_{12}\left[
\begin{array}{c}
                        m_{12}~~~m_{22}\\ m_{11}
\end{array}
\right]
& = &\left([l_{12}-l_{11}][l_{11}-l_{22}]\right)^{1/2}
\left[
\begin{array}{c}
                        m_{12}~~~m_{22}\\ m_{11}+1
\end{array}
\right],\\[2mm]
E_{21}\left[
\begin{array}{c}
                        m_{12}~~~m_{22}\\ m_{11}
\end{array}
\right]
& = &\left([l_{12}-l_{11}+1][l_{11}-l_{22}-1]\right)^{1/2}
\left[
\begin{array}{c}
                        m_{12}~~~m_{22}\\ m_{11}-1
\end{array}
\right],~~~~~~~(4.7)
\end{eqnarray*}
where $l_{ij}=m_{ij}-i$ for $i=1,2$ and $l_{ij}=m_{ij}-i+2$ for
$i=3,4$.\\

 On the other hands, $V_{0}^{q}$ is decomposed into the tensor
product $$V_{0}^{q}=V_{0,l}^{q}\otimes V_{0,r}^{q}.\eqno(4.8)$$
where $V_{0,l}^{q}$ and $V_{0,r}^{q}$ are a $U_{q}[gl(2)_{l}]$-
and a $U_{q}[gl(2)_{r}]$-fidirmods, respectively. Therefore, the GZ
basis of $V_{0}^{q}$ is the tensor product $$
\left[
\begin{array}{c}
                        m_{13}~~~m_{23}\\ m_{11}
\end{array}
\right]
\otimes
\left[
\begin{array}{c}
                        m_{33}~~~m_{43}\\ m_{31}
\end{array}
\right]
\equiv
\left[
\begin{array}{c}
                            [m]_{l}\\ m_{11}
\end{array}
\right]
\otimes
\left[
\begin{array}{c}
                            [m]_{r}\\ m_{31}
\end{array}
\right]
\equiv
(m)_{l}\otimes (m)_{r}
\equiv
(m)
\eqno(4.9)$$
between the GZ basis  of $V_{0,l}^{q}$ spanned on the vectors
$(m)_{l}$ and the GZ basis  of $V_{0,r}^{q}$ spanned on the
vectors $(m)_{r}$.  Following the approach of Ref. $^{20}$ and
keeping the notations used there, we can represent the GZ basis
(4.9) of $V_{0}^{q}$ in the form $$\left[
\begin{array}{lcr}

\begin{array}{c}
                        m_{13}~~~m_{23}\\ m_{11}
\end{array}
;
\begin{array}{c}
                        m_{33}~~~m_{43}\\ m_{31}
\end{array}
\end{array}
\right]
\equiv
\left[
\begin{array}{lcr}

\begin{array}{c}
                            [m]_{l}\\ m_{11}
\end{array}
;
\begin{array}{c}
                            [m]_{r}\\ m_{31}
\end{array}
\end{array}
\right]
\equiv
(m)
\eqno(4.10)$$
 Then, the highest weight $\Lambda$ is given by the first row
(signature) $[m_{13},m_{23},m_{33},m_{43}]\equiv
[[m]_{l},[m]_{r}]\equiv [m]$ common for all the GZ basis vectors
(4.10) of $V_{0}^{q}$: $$V_{0}^{q}\equiv
V_{0}^{q}(\Lambda)=V_{0}^{q}([m])= V_{0,l}^{q}([m]_{l})\otimes
V_{0,r}^{q}([m]_{r})\eqno(4.11)$$ The explicit action of
$U_{q}[gl(2/2)_{0}]$ on $V_{0}^{q}([m])$ follows directly from
(4.7) and : $$g_{0}(m)=g_{0,l}(m)_{l}\otimes (m)_{r} +
(m)_{l}\otimes g_{0,r}(m)_{r}\eqno(4.12)$$ for $g_{0}\equiv
g_{0,l}\oplus g_{0,r}\in U_{q}[gl(2/2)_{0}]$ and $(m)\in
V_{0}^{q}([m])$.\\

   The GZ basis vector $$
\left[
\begin{array}{c}
                        m_{13}~~~m_{23}\\ m_{13}
\end{array}
;
\begin{array}{c}
                        m_{33}~~~m_{43}\\ m_{33}
\end{array}
\right]
\equiv
\left[
\begin{array}{c}
                            [m]_{l}\\ m_{13}
\end{array}
;
\begin{array}{c}
                            [m]_{r}\\ m_{33}
\end{array}
\right]
\equiv (M)\eqno(4.13)$$
satisfying the conditions
\begin{tabbing}
\=123456789123456789123456\=$E_{ii}(M)$~\==~\=$m_{i3}(M)$
{}~~~~$i=1,2,3,4$123456789123.\=\kill
\>\>$E_{ii}(M)$\>=\>$m_{i3}(M)$,~~~~
$i=1,2,3,4$,\\[1mm]
\>\>$E_{12}(M)$\>=\>$E_{34}(M)~=~0$\>~~~(4.14)
\end{tabbing}
by definition, is the highest weight vector in $V_{0}^{q}([m])$.
Therefore, as in the classical case ($q=1$) the highest weight
$[m]$ is nothing but an ordered set of the eigen values of the
Cartan generators $E_{ii}$ on the highest weight vector $(M)$.
The latter is also highest weight vector in  $W^{q}([m])$ because
of the condition (4.1). All other, i.e.  lower weight, basis
vectors of $V_{0}^{q}$ can be obtained from the highest weight
vector $(M)$ through acting on the latter by monomials of
definite powers of the lowering generators $E_{21}$
and $E_{43}$:
\begin{eqnarray*}
{}~~~~~~~~(m)&=&\left({[m_{11}-m_{23}]![m_{31}-m_{43}]!\over
[m_{13}-m_{23}]![m_{13}-m_{11}]!
[m_{33}-m_{43}]![m_{33}-m_{31}]!}\right)^{1/2}\\[5mm]&&\times
(E_{21})^{m_{13}-m_{11}}(E_{43})^{m_{33}-m_{31}}(M),
{}~~~~~~~~~~~~~~~~~~~~~~~~~~~~~~~~~~~~~~~~~~~(4.15)
\end{eqnarray*}
where $[n]$'s are short hands of $${q^{2n}-q^{-2n}\over
q^{2}-q^{-2}}\equiv [n]_{q^{2}}\equiv [n],\eqno(4.16)$$ while
$$[n]!=[1][2]...[n-1][n].\eqno(4.17)$$

  Using (3.1-5) we can show that a q-analogue of the
Poincar\'e-Birkhoff-Witt theorem holds $^{26}$ (see also
$^{18}$), namely $U_{q}$ is a linear span of the elements of the
form $$g=(E_{31})^{\theta_{1}}(E_{32})^{\theta_{2}}
(E_{41})^{\theta_{3}}(E_{42})^{\theta_{4}}b,~~~b\in U_{q}(B),~~
\theta_{i}=0,1,~~i=1,2,3,4.\eqno(4.18)$$
Indeed, a right-ordered basis vector of $U_{q}$ like
$$h=(E_{23})^{\eta_{1}}(E_{24})^{\eta_{2}}
(E_{13})^{\eta_{3}}(E_{14})^{\eta_{4}}(E_{41})^{\theta_{1}}
(E_{31})^{\theta_{2}}(E_{42})^{\theta_{3}}(E_{32})^{\theta_{4}}h_{0},
$$
where $~\eta_{i},~ \theta_{i}=0,1,~ h_{0}\in U_{q}[gl(2/2)_{0}]$,
can be re-ordered and
expressed through the vectors (4.18) which are more
convenient for consulting the classical
case in Refs. $^{20,21}$.
Taking into account the fact that $V_{0}^{q}([m])$ is a
$U_{q}(B)$-module we have
$$W^{q}([m])=lin.env.\{(E_{31})^{\theta_{1}}(E_{32})^{\theta_{2}}
(E_{41})^{\theta_{3}}(E_{42})^{\theta_{4}}\otimes v\|~v\in
V_{0}^{q},~\theta_{1},...,\theta_{4}=0,1\}.\eqno(4.19)$$
Consequently, the vectors
$$\left|\theta_{1}\theta_{2},\theta_{3},\theta_{4};(m)\right>~:=
(E_{31})^{\theta_{1}}(E_{32})^{\theta_{2}}
(E_{41})^{\theta_{3}}(E_{42})^{\theta_{4}}\otimes
(m),\eqno(4.20)$$ all together span a basis of the module
$W^{q}([m])$. We shall call this basis induced in order to
distinguish it from the introduced later reduced basis and is
more convenient for us to investigate the reducibleness of
$W^{q}([m])$.\\

  The subspace $T^{q}$ consisting of
$$\left|\theta_{1},\theta_{2},\theta_{3},\theta_{4}\right>~:=
(E_{31})^{\theta_{1}}(E_{32})^{\theta_{2}}
(E_{41})^{\theta_{3}}(E_{42})^{\theta_{4}}\eqno(4.21)$$ can be
considered as
a $U_{q}[gl(2/2)_{0}]$-adjoint module (upto rescaling by $k_{i}$
in definite powers). The latter is
16-dimensional as begins from $\left|0,0,0,0\right>$ when
$\forall \theta_{i}=0$ and ends at $\left|1,1,1,1\right>$ when
$\forall
\theta_{i}=1$.
Therefore, $W^{q}([m])$, as a $U_{q}[gl(2/2)_{0}]$-module
$$W^{q}([m])=T^{q}\odot V_{0}^{q}([m]),\eqno(4.19')$$ is
reducible and can be decomposed into 16 finite-dimensional irreducible
$U_{q}[gl(2/2)_{0}]$-submodules $V_{k}^{q}([m]_{k})$,
$k=0,1,...,15$ :
$$W^{q}([m])=\bigoplus_{k=0}^{15}V_{k}^{q}([m]_{k}).\eqno(4.22)$$
Here, $[m]_{k}\equiv [m_{12},m_{22},m_{32},m_{42}]_{k}$ are the
local highest weights of the submodules $V_{k}^{q}$ in their GZ
bases denoted now as $$
\left[
\begin{array}{lcr}

\begin{array}{c}
                        m_{12}~~~m_{22}\\ m_{11}
\end{array}
;
\begin{array}{c}
                        m_{32}~~~m_{42}\\ m_{31}
\end{array}
\end{array}
\right]_{k}
\equiv
(m)_{k}.
\eqno(4.23)$$
The highest weight $[m]_{0}\equiv [m]$ of $V_{0}^{q}$ being also
the highest weight of $W^{q}$ is referred as a global highest
weight. We call $[m]_{k}$, $k\geq 1$ the local highest weights
in the sense that they characterize only the submodules
$V_{k}^{q}\subset W^{q}$ as $U_{q}[gl(2/2)_{0}]$-fidirmods,
while the global highest weight $[m]$ characterizes the
$U_{q}[gl(2/2)]$-module $W^{q}$ as the whole. In the same way
we define the local highest weight vectors $(M)_{k}$ in
$V_{k}^{q}$ as those $(m)_{k}$ satisfying the conditions (cf. (4.14))
\begin{eqnarray*}
{}~~~~~~~~~~~~~~~~~~~~~~~~~~~~E_{ii}(M)_{k}&=&m_{i2}(M)_{k},
{}~~~~i=1,2,3,4,\\[1mm]
                     E_{12}(M)_{k}&=&E_{34}(M)_{k}~=~0.
{}~~~~~~~~~~~~~~~~~~~~~~~~~~~~~~~~~~~(4.24)
\end{eqnarray*}
The highest weight vector $(M)$ of $V_{0}^{q}$ is also the global
highest weight vector in $W^{q}$  for which the condition (see
(4.1)) $$E_{23}(M)=0\eqno(4.25)$$ and  the conditions (4.24)
simultaneously hold.\\

  Let us denote by $\Gamma_{k}^{q}$ the basis system  spanned on
the basis vectors $(m)_{k}$ (4.23) in each $V_{k}^{q}([m])$.
For a basis of  $W^{q}$ we can choose the union
$\Gamma^{q}=\bigcup_{k=0}^{15}\Gamma_{k}^{q}$ of all the
bases $\Gamma_{k}^{q}$, namely, a basis vector of $W^{q}$ has
to be identified with one of the vectors $(m)_{k}$,
$k=0,1,...,15$.	The basis $\Gamma^{q}$ is referred as a
($U_{q}[gl(2/2)_{0}]$-) reduced basis. It is clear that every
basis $\Gamma_{k}^{q}=\Gamma_{k}([m]_{k})^{q}$ is labelled by a
local highest weight $[m]_{k}$, while the basis
$\Gamma^{q}=\Gamma^{q}([m])$ is labelled by the global
highest weight $[m]$. Going ahead, we modify the notation (4.23)
for the basis vectors in $\Gamma^{q}$ as follows (see (3.54) in
Ref. $^{20}$) $$
\left[
\begin{array}{lccc}

                        m_{13}& m_{23}&  m_{33}& m_{43} \\

                        m_{12}& m_{22}&  m_{32}&  m_{42}\\
m_{11}&   0   &  m_{31}&    0
\end{array}
\right]_{k}
\equiv
\left[
\begin{array}{lcr}

\begin{array}{c}
                        m_{12}~~~m_{22}\\ m_{11}
\end{array}
;
\begin{array}{c}
                        m_{32}~~~m_{42}\\ m_{31}
\end{array}
\end{array}
\right]_{k}
\equiv
(m)_{k},
\eqno(4.26)$$
with $k$ running from 0 to 15 as for $k=0$ we have to take into
account $m_{i2}=m_{i3}$, $i=1,2,3,4$, i.e.
$$ (m)_{0}\equiv (m)=
\left[
\begin{array}{lccc}

                        m_{13}& m_{23}&  m_{33}& m_{43} \\

                        m_{13}& m_{23}&  m_{33}&  m_{43}\\
m_{11}&   0   &  m_{31}&    0
\end{array}
\right].
\eqno(4.27)
$$ In (4.26) the first row $[m]=[m_{13},m_{23},m_{33},m_{43}]$
being the (global) highest weight of $W^{q}$ is fixed for all the
vectors in the whole $W^{q}$ and characterizes the module itself,
while the second row is a (local) highest weight of some
submodule $V_{k}^{q}$ and tells us that the considered basis
vector $(m)_{k}$ of $W^{q}$ belongs to this submodule in the
decomposition (4.22) corresponding to the branching rule
$U_{q}[gl(2/2)]\supset U_{q}[gl(2/2)_{0}]$.\\

   It is easy to  see that the highest vectors $(M)_{k}$  in the
notation (4.26) are $$ (M)_{k}=
\left[
\begin{array}{lccc}

                        m_{13}& m_{23}&  m_{33}& m_{43} \\

                        m_{12}& m_{22}&  m_{32}&  m_{42}\\
m_{12}&   0   &  m_{32}&    0
\end{array}
\right]_{k},~~~k=0,1,...15.
\eqno(4.28)$$
The (global) highest weight vector $(M)$ (4.13) is given now by
$$ (M)=
\left[
\begin{array}{lccc}

                        m_{13}& m_{23}&  m_{33}& m_{43} \\

                        m_{13}& m_{23}&  m_{33}&  m_{43}\\
m_{13}&   0   & m_{33}& 0 \end{array} \right].  \eqno(4.29)$$ Let
us denote by $(m)_{k}^{\pm ij}$ a GZ vector obtained from
$(m)_{k}$ by replacing the element $m_{ij}$ of the latter by
$m_{ij}\pm 1$.  We can prove that the highest weight vectors
$(M)_{k}$ expressed in terms of the induced basis (4.20) have the
following explicit forms
\begin{eqnarray*}
{}~~~~~~(M)_{0}& = &a_{0}\left|0,0,0,0;(M)\right>,~~~~a_{0}\equiv
1,\\[2mm] (M)_{1}& = &a_{1}\left|0,1,0,0;(M)\right>,\\[2mm]
(M)_{2}& = &a_{2}\left\{\left|1,0,0,0;(M)\right>
+q^{4l}[2l]^{-1/2}
\left|0,1,0,0;(M)^{-11}\right>\right\},\\[2mm]
(M)_{3}& = &a_{3}\left\{\left|0,0,0,1;(M)\right>
-q^{-4l'-2}[2l']^{-1/2}
\left|0,1,0,0;(M)^{-31}\right>\right\},\\[2mm]
(M)_{4}& =
&a_{4}\left\{\left|0,0,1,0;(M)\right>
+q^{4l}[2l]^{-1/2}\left|0,0,0,1;(M)^{-11}\right>\right.\\[2mm]
& &-q^{-4l'-2}[2l']^{-1/2}
\left|1,0,0,0;(M)^{-31}\right>\\[2mm]
& &\left.-q^{4l-4l'-2}\left([2l][2l']\right)^{-1/2}
\left|0,1,0,0;(M)^{-11-31}\right>\right\},\\[2mm]
(M)_{5}& = &a_{5}\left|0,1,0,1;(M)\right>,\\[4mm] (M)_{6}& =
&a_{6}\left\{\left|1,0,0,1;(M)\right>
+q^{-2}\left|0,1,1,0;(M)\right>\right.\\[2mm]
& &\left.
+q^{4l}(q^{2}+q^{-2})[2l]^{-1/2}
\left|0,1,0,1;(M)^{-11}\right>\right\},\\[2mm]
(M)_{7}& = &a_{7}\left\{\left|1,0,1,0;(M)\right>+
q^{4l}[2l]^{-1/2}\left|1,0,0,1;(M)^{-11}\right>\right.\\[2mm]
& &+q^{4l-2}[2l]^{-1/2}\left|0,1,1,0;(M)^{-11}\right>\\[2mm]
& &\left.
+q^{8l-4}\left(q^{2}+q^{-2}\right)^{1/2}\left([2l][2l-1]\right)^{-1/2}
\left|0,1,0,1;(M)^{-11-11}\right>\right\},\\[2mm]
(M)_{8}& =
&a_{8}\left|1,1,0,0;(M)\right>,\\[2mm]
(M)_{9}& =
&a_{9}\left\{\left|1,0,0,1;(M)\right>
-q^{2}\left|0,1,1,0;(M)\right>\right.\\[2mm] & &\left.
-q^{-4l'}(q^{2}+q^{-2})[2l']^{-1/2}
\left|1,1,0,0;(M)^{-31}\right>\right\},\\[2mm]
(M)_{10}& =
&a_{10}\left\{\left|0,0,1,1;(M)\right>
-q^{-4l'-4}[2l']^{-1/2}
\left|1,0,0,1;(M)^{-31}\right>\right.\\[2mm]
& &+q^{-4l'-2}[2l']^{-1/2}
\left|0,1,1,0;(M)^{-31}\right>\\[2mm] &
&\left. +q^{-8l'}\left(q^{2}+q^{-2}\right)^{1/2}
\left([2l'][2l'-1]\right)^{-1/2}
\left|1,1,0,0;(M)^{-31-31}\right>\right\},\\[2mm]
(M)_{11}& =
&a_{11}\left|1,1,0,1;(M)\right>,\\[2mm]
(M)_{12}& = &a_{12}\left\{\left|1,1,1,0;(M)\right>
+q^{4l}[2l]^{-1/2}
\left|1,1,0,1;(M)^{-11}\right>\right\},\\[2mm]
(M)_{13}& = &a_{13}\left\{\left|0,1,1,1;(M)\right>
+q^{-4l'-2}[2l']^{-1/2}
\left|1,1,0,1;(M)^{-31}\right>\right\},\\[2mm]
(M)_{14}& = &a_{14}\left\{\left|1,0,1,1;(M)\right>
+q^{4l}[2l]^{-1/2}
\left|0,1,1,1;(M)^{-11}\right>\right.\\[2mm] &
&+q^{-4l'-2}[2l']^{-1/2}
\left|1,1,1,0;(M)^{-31}\right>\\[2mm] &
&\left.
+q^{4l-4l'-2}\left([2l][2l']\right)^{-1/2}
\left|1,1,0,1;(M)^{-11-31}\right>\right\},\\[2mm]
(M)_{15}& =
&a_{15}\left|1,1,1,1;(M)\right>,
{}~~~~~~~~~~~~~~~~~~~~~~~~~~~~~~~~~~~~~~~~~~~~~~~~~~~~~~(4.30)
\end{eqnarray*}\\ where
$l={1\over 2}(m_{13}-m_{23})$ and $l'={1\over 2}(m_{33}-m_{43})$,
while $a_{k}=a_{k}(q)$ are coefficients depending on $q$. Indeed,
$(M)_{k}$ given in (4.30) form a set of all linear independent
vectors satisfying the conditions (4.24).
For a further convenience, let us rescale the coefficients $a_{k}$ as
follows
\begin{eqnarray*}
{}~~~~~~~~a_{0}& = &c_{0}\equiv 1~,~~~~~~~~~~~~~~~~~~~~~~~~~~~~~~~
a_{8}~~=~~q^{-2}c_{8}~,\\[1mm] a_{1}& = &-q^{-2}c_{1}~,
{}~~~~~~~~~~~~~~~~~~~~~~~~~~~~~~ a_{9}~~=~~-q^{-2}\left({[2l']\over
[2][2l'+2]}\right)^{1/2}c_{9}~,\\[1mm] a_{2}& =
&-q^{-2}\left({[2l]\over [2l+1]}\right)^{1/2}c_{2}~,~~~~~~~~~~
a_{10}~~=~~q^{2}\left({[2l'-1]\over[2l'+1]}\right)^{1/2}c_{10}~,
\\[1mm]
a_{3}& =
&\left({[2l']\over[2l'+1]}\right)^{1/2}c_{3}~,~~~~~~~~~~~~~~~~~
a_{11}~~=~~q^{-2}c_{11}~,\\[1mm] a_{4}& =
&\left({[2l][2l']\over[2l+1][2l'+1]}\right)^{1/2}c_{4}~,~~~~~~~~
a_{12}~~=~~q^{-2}\left({[2l]\over[2l+1]}\right)^{1/2}c_{12}~,\\[1mm]
a_{5}& = &q^{-2}c_{5}~,~~~~~~~~~~~~~~~~~~~~~~~~~~~~~~~~
a_{13}~~=~~\left({[2l']\over[2l'+1]}\right)^{1/2}c_{13}~,\\[1mm]
a_{6}& = &q^{-2}\left({[2l]\over
[2][2l+2]}\right)^{1/2}c_{6}~,~~~~~~~~~~
a_{14}~~=~~\left({[2l][2l']\over[2l+1][2l'+1]}\right)^{1/2}c_{14}~,\\[1mm]
a_{7}& =
&q^{-2}\left({[2l-1]\over[2l+1]}\right)^{1/2}c_{7},~~~~~~~~~~~~~~
a_{15}~~=~~c_{15}~, ~~~~~~~~~~~~~~~~~~~~~~~~~(4.31)
\end{eqnarray*} where $c_{k}=c_{k}(q)$ are some other
constants which may depend on $q$. Looking at (4.30) we  easily
identify the highest weights $[m]_{k}$
\begin{tabbing} \=12345679123456789\= $[m]_{kk}$ \= =x \=
$[m_{13}-1,m_{23}-1,m_{33}+1,m_{43}+1]$,\=\kill
\>\>
$[m]_{0}$ \> = \> $[m_{13}, m_{23}, m_{33}, m_{43}]$,\\[2mm]
\>\>$[m]_{1}$ \> = \> $[m_{13}, m_{23}-1, m_{33}+1,
m_{43}]$,\\[2mm] \>\>$[m]_{2}$ \> = \> $[m_{13}-1, m_{23},
m_{33}+1, m_{43}]$,\\[2mm] \>\>$[m]_{3}$ \> = \> $[m_{13},
m_{23}-1, m_{33}, m_{43}+1]$,\\[2mm] \>\>$[m]_{4}$ \> = \>
$[m_{13}-1, m_{23}, m_{33}, m_{43}+1]$,\\[2mm]
\>\>$[m]_{5}$ \> =
\> $[m_{13},m_{23}-2,m_{33}+1,m_{43}+1]$,\\[2mm]
\>\>$[m]_{6}$ \>
= \> $[m_{13}-1,m_{23}-1,m_{33}+1,m_{43}+1]_{6}$,\\[2mm]
\>\>$[m]_{7}$ \> = \>
$[m_{13}-2,m_{23},m_{33}+1,m_{43}+1]$,\\[2mm]
\>\>$[m]_{8}$ \> =\> $[m_{13}-1, m_{23}-1, m_{33}+2, m_{43}]$,
\\[2mm]
\>\>$[m]_{9}$\> = \> $[m_{13}-1, m_{23}-1, m_{33}+1, m_{43}+1]_{9}$,
\\[2mm]
\>\>$[m]_{10}$ \> = \>$[m_{13}-1,m_{23}-1,m_{33},m_{43}+2]$,\\[2mm]
\>\>$[m]_{11}$ \> =
\> $[m_{13}-1,m_{23}-2,m_{33}+2,m_{43}+1]$,\\[2mm] \>\>$[m]_{12}$
\> = \> $[m_{13}-2,m_{23}-1,m_{33}+2,m_{43}+1]$,\\[2mm]
\>\>$[m]_{13}$ \> = \>
$[m_{13}-1,m_{23}-2,m_{33}+1,m_{43}+2]$,\\[2mm] \>\>$[m]_{14}$ \>
= \> $[m_{13}-2,m_{23}-1,m_{33}+1,m_{43}+2]$,\\[2mm]
\>\>$[m]_{15}$ \> = \>
$[m_{13}-2,m_{23}-2,m_{33}+2,m_{43}+2]$.~~~~~~~~~~~~~~~(4.32)
\end{tabbing}
\vspace*{2mm}
In the latest formula (4.32), with the exception of
$[m]_{6}$ and $[m]_{9}$ where a degeneration is present, we skip
the subscript $k$ in the r.h.s.. The proofs of (4.30) and (4.32)
follow from direct computations.\\

   Using the rule (4.15) which now reads
\begin{eqnarray*}
{}~~~~~~~~~(m)_{k}&=&\left({[m_{11}-m_{22}]![m_{31}-m_{42}]!\over
[m_{12}-m_{22}]![m_{12}-m_{11}]!
[m_{32}-m_{42}]![m_{32}-m_{31}]!}\right)^{1/2}\\[5mm]&&\times
(E_{21})^{m_{12}-m_{11}}(E_{43})^{m_{32}-m_{31}}(M)_{k}
{}~~~~~~~~~~~~~~~~~~~~~~~~~~~~~~~~~~~~~~~(4.15')
\end{eqnarray*}\\[2mm]
we can find all the basis vectors $(m)_{k}$ :
\begin{eqnarray*}
(m)_{0}& = &\left|0,0,0,0;(m)\right>,\\[4mm]
(m)_{1}
& = &c_{1}\left\{q^{2(l'-p')}\left(
{[l_{13}-l_{11}][l_{31}-l_{43}-1] \over
[2l+1][2l'+1]}\right)^{1/2}\left|1,0,0,0;(m)^{+11-31}\right>\right.
\\ &   &-q^{2(-l+p+l'-p')}\left(
{[l_{11}-l_{23}][l_{31}-l_{43}-1]\over
[2l+1][2l'+1]}\right)^{1/2}\left|0,1,0,0;(m)^{-31}\right>\\ &
&+\left( {[l_{13}-l_{11}][l_{33}-l_{31}+1]\over
[2l+1][2l'+1]}\right)^{1/2}\left|0,0,1,0;(m)^{+11}\right>\\ &
&-\left. q^{2(-l+p)}\left(
{[l_{11}-l_{23}][l_{33}-l_{31}+1]\over
[2l+1][2l'+1]}\right)^{1/2}\left|0,0,0,1;(m)\right>\right\},
\\[4mm]
(m)_{2}
& = &c_{2}\left\{-q^{2(l'-p')}\left(
{[l_{11}-l_{23}][l_{31}-l_{43}-1] \over
[2l+1][2l'+1]}\right)^{1/2}\left|1,0,0,0;(m)^{+11-31}\right>\right.
\\ &   &-q^{2(l+p+l'-p'+1)}\left(
{[l_{13}-l_{11}][l_{31}-l_{43}-1]\over
[2l+1][2l'+1]}\right)^{1/2}\left|0,1,0,0;(m)^{-31}\right>\\ &
&-\left( {[l_{11}-l_{23}][l_{33}-l_{31}+1]\over
[2l+1][2l'+1]}\right)^{1/2}\left|0,0,1,0;(m)^{+11}\right>\\ &
&-\left. q^{2(l+p+1)}\left(
{[l_{13}-l_{11}][l_{33}-l_{31}+1]\over
[2l+1][2l'+1]}\right)^{1/2}\left|0,0,0,1;(m)\right>\right\},
\\[4mm]
(m)_{3}& = &c_{3}\left\{q^{-2(l'+p'+1)}\left(
{[l_{13}-l_{11}][l_{33}-l_{31}+1] \over
[2l+1][2l'+1]}\right)^{1/2}\left|1,0,0,0;(m)^{+11-31}\right>\right.
\\ &   &-q^{-2(l-p+l'+p'+1)}\left(
{[l_{11}-l_{23}][l_{33}-l_{31}+1]\over
[2l+1][2l'+1]}\right)^{1/2}\left|0,1,0,0;(m)^{-31}\right>\\ &
&-\left( {[l_{13}-l_{11}][l_{31}-l_{43}-1]\over
[2l+1][2l'+1]}\right)^{1/2}\left|0,0,1,0;(m)^{+11}\right>\\ &
&\left. +q^{2(-l+p)}\left(
{[l_{11}-l_{23}][l_{31}-l_{43}-1]\over
[2l+1][2l'+1]}\right)^{1/2}\left|0,0,0,1;(m)\right>\right\},
\\[4mm]
(m)_{4}& = &c_{4}\left\{-q^{-2(l'+p'+1)}\left(
{[l_{11}-l_{23}][l_{33}-l_{31}+1] \over
[2l+1][2l'+1]}\right)^{1/2}\left|1,0,0,0;(m)^{+11-31}\right>\right.
\\ &   &-q^{2(l+p-l'-p')}\left(
{[l_{13}-l_{11}][l_{33}-l_{31}+1]\over
[2l+1][2l'+1]}\right)^{1/2}\left|0,1,0,0;(m)^{-31}\right>\\ &
&+\left( {[l_{11}-l_{23}][l_{31}-l_{43}-1]\over
[2l+1][2l'+1]}\right)^{1/2}\left|0,0,1,0;(m)^{+11}\right>\\
& &\left. +q^{2(l+p+1)}\left(
{[l_{13}-l_{11}][l_{31}-l_{43}-1]\over
[2l+1][2l'+1]}\right)^{1/2}\left|0,0,0,1;(m)\right>\right\},
\\[4mm]
(m)_{5}& = &c_{5}\left\{q^{-2}\left(
{[l_{13}-l_{11}][l_{13}-l_{11}-1] \over
[2l+1][2l+2]}\right)^{1/2}\left|1,0,1,0;(m)^{+11+11-31}\right>\right.
\\ &   &-q^{2(-l+p)}\left( {[l_{13}-l_{11}][l_{11}-l_{23}+1]
\over
[2l+1][2l+2]}\right)^{1/2}\left|1,0,0,1;(m)^{+11-31}\right>\\ &
&-q^{2(-l+p-1)}\left( {[l_{13}-l_{11}][l_{11}-l_{23}+1] \over
[2l+1][2l+2]}\right)^{1/2}\left|0,1,1,0;(m)^{+11-31}\right>\\ &
&\left. +q^{2(-2l+2p-1)}\left( {[l_{11}-l_{23}][l_{11}-l_{23}+1]
\over
[2l+1][2l+2]}\right)^{1/2}\left|0,1,0,1;(m)^{-31}\right>\right\},
\\[4mm]
(m)_{6}
& = &c_{6}\left\{-q^{-2}(q^{2}+q^{-2})
\left({[l_{13}-l_{11}-1][l_{11}-l_{23}+1]\over
[2][2l][2l+2]}\right)^{1/2}\left|1,0,1,0;(m)^{+11+11-31}
\right>\right.\\
&   &+q^{2(-l+p)}{([l_{11}-l_{23}]-q^{4(l+1)}
[l_{13}-l_{11}-1])\over \left([2][2l][2l+2]\right)^{1/2}}
\left|1,0,0,1;(m)^{+11-31}\right>\\
& &+q^{2(-l+p-1)}{([l_{11}-l_{23}]-q^{4(l+1)}
[l_{13}-l_{11}-1])\over
\left([2][2l][2l+2]\right)^{1/2}}
\left|0,1,1,0;(m)^{+11-31}\right>\\
&   &\left. +(q^{2}+q^{-2})q^{2(2p+1)}\left({[l_{11}-l_{23}]
[l_{13}-l_{11}]\over
[2][2l][2l+2]}\right)^{1/2}\left|0,1,0,1;(m)^{-31}
\right>\right\},\\[4mm]
(m)_{7}
& = &c_{7}\left\{q^{-2}\left(
{[l_{11}-l_{23}][l_{11}-l_{23}+1] \over
[2l][2l+1]}\right)^{1/2}\left|1,0,1,0;(m)^{+11+11-31}
\right>\right.\\
& &+q^{2(l+p+1)}\left( {[l_{11}-l_{23}][l_{13}-l_{11}-1]
\over [2l][2l+1]}\right)^{1/2}\left|1,0,0,1;(m)^{+11-31}
\right>\\
&  &+q^{2(l+p)}\left( {[l_{11}-l_{23}][l_{13}-l_{11}-1] \over
[2l][2l+1]}\right)^{1/2}\left|0,1,1,0;(m)^{+11-31}
\right>\\
& &\left. +q^{2(2l+2p+1)}\left( {[l_{13}-l_{11}]
[l_{13}-l_{11}-1]\over
[2l][2l+1]}\right)^{1/2}\left|0,1,0,1;(m)^{-31}
\right>\right\},\\[4mm]
(m)_{8}
& = &c_{8}\left\{q^{2}\left(
{[l_{33}-l_{31}+1][l_{33}-l_{31}+2] \over
[2l'+1][2l'+2]}\right)^{1/2}\left|0,0,1,1;(m)^{+11}\right>
\right.\\
       &   &+q^{2(l'-p')}\left(
{[l_{33}-l_{31}+2][l_{31}-l_{43}-1] \over
[2l'+1][2l'+2]}\right)^{1/2}\left|1,0,0,1;(m)^{+11-31}\right>\\
& &-q^{2(l'-p'+1)}\left( {[l_{33}-l_{31}+2][l_{31}-l_{43}-1]
\over
[2l'+1][2l'+2]}\right)^{1/2}\left|0,1,1,0;(m)^{+11-31}\right>\\
& &\left. +q^{2(2l'-2p'+3)}\left( {[l_{31}-l_{43}-2]
[l_{31}-l_{43}-1]\over
[2l'+1][2l'+2]}\right)^{1/2}\left|1,1,0,0;(m)^{+11-31-31}
\right>\right\},\\[4mm]
(m)_{9}
& = &c_{9}\left\{-q^{2}(q^{2}+q^{-2})\left(
{[l_{33}-l_{31}+1][l_{31}-l_{43}-1] \over
[2][2l'][2l'+2]}\right)^{1/2}\left|0,0,1,1;(m)^{+11}
\right>\right.\\
& &-q^{2(l'-p')}{([l_{31}-l_{43}-2]-q^{-4(l'+1)}
[l_{33}-l_{31}+1])\over
\left([2][2l'][2l'+2]\right)^{1/2}}\left|1,0,0,1;(m)^{+11-31}
\right>\\
& &+q^{2(l'-p'+1)}{([l_{31}-l_{43}-2]-q^{-4(l'+1)}
[l_{33}-l_{31}+1])\over
\left([2][2l'][2l'+2]\right)^{1/2}}\left|0,1,1,0;(m)^{+11-31}
\right>\\
&  &\left. +q^{2(-2p'+1)}(q^{2}+q^{-2})\left(
{[l_{31}-l_{43}-2][l_{33}-l_{31}+2] \over
[2][2l'][2l'+2]}\right)^{1/2}\left|1,1,0,0;(m)^{+11-31-31}
\right>\right\},\\[4mm]
(m)_{10}
& = &c_{10}\left\{q^{2}\left(
{[l_{31}-l_{43}-2][l_{31}-l_{43}-1] \over
[2l'][2l'+1]}\right)^{1/2}\left|0,0,1,1;(m)^{+11}
\right>\right.\\
& &-q^{-2(l'+p'+1)}\left( {[l_{33}-l_{31}+1][l_{31}-l_{43}-2]
\over [2l'][2l'+1]}\right)^{1/2}\left|1,0,0,1;(m)^{+11-31}
\right>\\
& &+q^{-2(l'+p')}\left( {[l_{33}-l_{31}+1][l_{31}-l_{43}-2]
\over [2l'][2l'+1]}\right)^{1/2}\left|0,1,1,0;(m)^{+11-31}
\right>\\ & &\left. +q^{-2(2l'+2p'-1)}\left(
{[l_{33}-l_{31}+1][l_{33}-l_{31}+2] \over
[2l'][2l'+1]}\right)^{1/2}\left|1,1,0,0;(m)^{+11-31-31}
\right>\right\},\\[4mm]
(m)_{11} & = &c_{11}\left\{\left(
{[l_{13}-l_{11}-1][l_{33}-l_{31}+2] \over
[2l+1][2l'+1]}\right)^{1/2}\left|1,0,1,1;(m)^{+11+11-31}
\right>\right.\\
&  &-q^{2(-l+p+1)}\left( {[l_{11}-l_{23}+1][l_{33}-l_{31}+2]
\over
[2l+1][2l'+1]}\right)^{1/2}\left|0,1,1,1;(m)^{+11-31}
\right>\\
& &-q^{2(l'-p'+1)}\left( {[l_{13}-l_{11}-1][l_{31}-l_{43}-2]
\over [2l+1][2l'+1]}\right)^{1/2}\left|1,1,1,0;(m)^{+11+11-31-31}
\right>\\
& &\left. +q^{2(-l+p+l'-p'+2)}\left(
{[l_{11}-l_{23}+1][l_{31}-l_{43}-2] \over
[2l+1][2l'+1]}\right)^{1/2}\left|1,1,0,1;(m)^{+11-31-31}
\right>\right\},\\[4mm]
(m)_{12}& = &c_{12}\left\{-\left(
{[l_{11}-l_{23}+1][l_{33}-l_{31}+2] \over
[2l+1][2l'+1]}\right)^{1/2}\left|1,0,1,1;(m)^{+11+11-31}
\right>\right.\\
&  &-q^{2(l+p+2)}\left( {[l_{13}-l_{11}-1][l_{33}-l_{31}+2]
\over [2l+1][2l'+1]}\right)^{1/2}\left|0,1,1,1;(m)^{+11-31}
\right>\\
& &+q^{2(l'-p'+1)}\left( {[l_{11}-l_{23}+1][l_{31}-l_{43}-2]
\over [2l+1][2l'+1]}\right)^{1/2}\left|1,1,1,0;(m)^{+11+11-31-31}
\right>\\
& &\left. +q^{2(l+p+l'-p'+3)}\left(
{[l_{13}-l_{11}-1][l_{31}-l_{43}-2] \over
[2l+1][2l'+1]}\right)^{1/2}\left|1,1,0,1;(m)^{+11-31-31}
\right>\right\},
\end{eqnarray*}
\begin{eqnarray*}
(m)_{13}
& = &c_{13}\left\{-\left(
{[l_{13}-l_{11}-1][l_{31}-l_{43}-2] \over
[2l+1][2l'+1]}\right)^{1/2}\left|1,0,1,1;(m)^{+11+11-31}
\right>\right.\\
& &+q^{2(-l+p+1)}\left( {[l_{11}-l_{23}+1][l_{31}-l_{43}-2]
\over [2l+1][2l'+1]}\right)^{1/2}\left|0,1,1,1;(m)^{+11-31}
\right>\\
& &-q^{-2(l'+p')}\left( {[l_{13}-l_{11}-1][l_{33}-l_{31}+2]
\over [2l+1][2l'+1]}\right)^{1/2}\left|1,1,1,0;(m)^{+11+11-31-31}
\right>\\
&   &\left. +q^{2(-l+p-l'-p'+1)}\left(
{[l_{11}-l_{23}+1][l_{33}-l_{31}+2] \over
[2l+1][2l'+1]}\right)^{1/2}\left|1,1,0,1;(m)^{+11-31-31}
\right>\right\},\\[4mm]
(m)_{14} & = &c_{14}\left\{\left(
{[l_{11}-l_{23}+1][l_{31}-l_{43}-2] \over
[2l+1][2l'+1]}\right)^{1/2}\left|1,0,1,1;(m)^{+11+11-31}
\right>\right.\\
&  &+q^{2(l+p+2)}\left( {[l_{13}-l_{11}-1][l_{31}-l_{43}-2]
\over [2l+1][2l'+1]}\right)^{1/2}\left|0,1,1,1;(m)^{+11-31}
\right>\\
& &+q^{-2(l'+p')}\left( {[l_{11}-l_{23}+1][l_{33}-l_{31}+2]
\over [2l+1][2l'+1]}\right)^{1/2}\left|1,1,1,0;(m)^{+11+11-31-31}
\right>\\
&  &\left. +q^{2(l+p-l'-p'+2)}\left(
{[l_{13}-l_{11}-1][l_{33}-l_{31}+2] \over
[2l+1][2l'+1]}\right)^{1/2}\left|1,1,0,1;(m)^{+11-31-31}
\right>\right\},\\[4mm]
(m)_{15}&=& c_{15}\left|1,1,1,1;(m)\right>,
{}~~~~~~~~~~~~~~~~~~~~~~~~~~~~~~~~~~~~~~~~~~~~~~~~~~~~~~~~~~~~~~(4.33)
\end{eqnarray*}
\\
where $l={1\over 2}(m_{13}-m_{23})$, $p=m_{11}-{1\over
2}(m_{13}+m_{23})$, $l'={1\over 2}(m_{33}-m_{43})$  and
$p'=m_{31}-{1\over 2}(m_{33}+m_{43})$.  The latest formula
(4.33), in fact, represents a way in which the reduced basis is
expressed in terms of the induced basis and vas versa it is not
a problem for us  to   find the invert relation between these
bases (see the Appendix).\\

  Taking into account all results obtained above we have proved
the following assertion
\\[2mm]
{\it Proposition} 1 : The $U_{q}[gl(2/2)]$-module $W^{q}$ is
decomposed as a direct sum (4.22) of sixteen $U_{q}[gl(2/2)_{0}]$-fidirmods
$V^{q}_{k}$, $k=0,1,...,15$, every one of which is characterized
by a highest weight $[m]_{k}$ given in (4.32) and is spanned by a
GZ basis $(m)_{k}$ given in (4.33).\\

   The decomposition (4.22) of $W^{q}([m])\equiv
W^{q}([m_{13}, m_{23}, m_{33}, m_{43}])$  can be re-written in
the form $^{21}$\\
\begin{eqnarray*}
W^{q}([m])&=&
V_{(00)}^{q}([m_{13}, m_{23}, m_{33}, m_{43}])\\[2mm]
& &\bigoplus_{i=0}^{min(1,2l)}
{}~\bigoplus_{j=0}^{min(1,2l')}
V_{(10)}^{q}([m_{13}-i, m_{23}+i-1, m_{33}-j+1, m_{43}+j])\\[2mm]
& &\bigoplus_{i=0}^{min(2,2l)}
V_{(11)}^{q}([m_{13}-i, m_{23}+i-2, m_{33}+1, m_{43}+1])\\[2mm]
& &\bigoplus_{j=0}^{min(2,2l')}
V_{(20)}^{q}([m_{13}-1, m_{23}-1, m_{33}-j+2, m_{43}+j])\\[2mm]
& &\bigoplus_{i=0}^{min(1,2l)}~\bigoplus_{j=0}^{min(1,2l')}
V_{(21)}^{q}([m_{13}-i-1, m_{23}+i-2, m_{33}-j+2, m_{43}+j+1])\\[2mm]
& &\bigoplus V_{(22)}^{q}([m_{13}-2, m_{23}-2, m_{33}+2, m_{43}+2])
{}~~~~~~~~~~~~~~~~~~~~~~~~(4.34)
\end{eqnarray*}
\\
where $V_{(ab)}^{q}([m]_{(ab)})$ is an alternative notation of
that submodule $V^{q}_{k}([m]_{k})$ with $[m]_{k} = [m]_{(ab)}$:
\begin{eqnarray*}
{}~~~~~~~~~~~~~~~~~~V_{(00)}^{q}([m]_{(00)})&\equiv &  V_{(00)}^{q}([m])
{}~ = V_{0}^{q}([m]),\\[1mm]
V_{(10)}^{q}([m]_{(10)})&\equiv & V_{(10)}^{q}([m]_{k}) =
V_{k}^{q}([m]_{k}), ~~ 1\leq k\leq 4,\\[1mm]
V_{(11)}^{q}([m]_{(11)})&\equiv & V_{(11)}^{q}([m]_{k}) =
V_{k}^{q}([m]_{k}), ~~ 5\leq k\leq 7,\\[1mm]
V_{(20)}^{q}([m]_{(20)})&\equiv & V_{(20)}^{q}([m]_{k}) =
V_{k}^{q}([m]_{k}), ~~ 8\leq k\leq 10,\\[1mm]
V_{(21)}^{q}([m]_{(21)})&\equiv & V_{(21)}^{q}([m]_{k}) =
V_{k}^{q}([m]_{k}), ~~ 11\leq k\leq 14,\\[1mm]
V_{(22)}^{q}([m]_{(22)})&\equiv & V_{(22)}^{q}([m]_{15}) =
V_{15}^{q}([m]_{15}).~~~~~~~~~~~~~~~~~~~~~~~~~~~~~(4.35)
\end{eqnarray*}
In a natural way we denote by $(m)_{(ab)}$ the GZ basis of
$V_{(ab)}^{q}([m]_{(ab)})$.
Thus, the basis of $W^{q}([m])$ is spanned on the set of all possible
patterns $^{21}$
$$(m)_{(ab)}\equiv \left[
\begin{array}{lccc}

                        m_{13}& m_{23}&  m_{33}& m_{43} \\

                        m_{12}& m_{22}&  m_{32}&  m_{42}\\
m_{11}&   0   &  m_{31}&    0
\end{array}
\right]_{(ab)},~~~~a,b\in \{0,1,2\}\eqno(4.36)$$
such that
\begin{eqnarray*}
{}~~~~~~~~~~~~~~~~~~~~~~~m_{12} &=& m_{13}-r-\theta (a-2) -\theta (b-2)
+1,\\[1mm]
         m_{22} &=& m_{23}+r-\theta (a-1) -\theta (b-1) -1,\\[1mm]
	    m_{32} &=& m_{33}+a-s+1,\\[1mm]
         m_{42} &=& m_{43}+b+s-1,
{}~~~~~~~~~~~~~~~~~~~~~~~~~~~~~~~~~~~~~~~~~~~~~(4.37)
\end{eqnarray*}
where
\begin{eqnarray*}
{}~~~~~~~~~~~~~~~~~~~~~~~~~~r &=& 1,...,1+min(a-b,2l'),\\[1mm]
s &=& 1,...,1+min(\left<a\right>+\left<b\right>,2l),
{}~~~~~~~~~~~~~~~~~~~~~~~~~~~~~(4.38)
\end{eqnarray*}

$$\theta (x)=\left\{
\begin{array}{ll}
1 & ~~~~ if ~~ x\geq 0,\\
0 & ~~~~ if ~~ x < 0
\end{array}\right.
\eqno(4.39)$$
\vspace*{1mm}
and

$$\left<i\right>=\left\{
\begin{array}{ll}
1 & ~~~~ for~~ odd~~ i,\\
0 & ~~~~ for~~ even~~ i~.
\end{array}\right.
\eqno(4.40)$$
\vspace*{2mm}

{\bf 4.2. Typical representations}\\

  The $U_{q}[gl(2/2)]$-module $W^{q}$ constructed is either
irreducible or indecomposable. We can verify that\\[2mm] {\it
Proposition} 2 : The induced $U_{q}[gl(2/2)]$-module $W^{q}$ is
irreducible if and only if the following condition holds
$$[l_{13}+l_{33}+3][l_{13}+l_{43}+3][l_{23}+l_{33}+3][l_{23}+l_{43}+3]
\neq 0.\eqno(4.41)$$ In this case we say that the module $W^{q}$
is typical, otherwise it is called atypical.\\

  The proof of the
latter proposition follows the one for the classical case
considered in Ref. $^{20}$ (cf. Ref. $^{18}$).  Indeed, by the
same argument we can conclude that $W^{q}$ is irreducible if and
only if
$$E_{24}E_{14}E_{23}E_{13}E_{31}E_{32}E_{41}E_{42}\otimes (M)
\neq 0.\eqno(4.42)$$
The latest condition (4.42) in turn can be
proved, after some elementary calculations, to be equivalent to
$$[E_{11}+E_{33}+1][E_{11}+E_{44}][E_{22}+E_{33}][E_{22}+E_{44}-1](M)
\neq 0,\eqno(4.42')$$
\vspace*{1mm}
which is nothing but the condition (4.41).\\

  Since $U_{q}[gl(2/2)]$ is generated by the even generators and
the odd Chevalley generators $E_{23}$ and $E_{32}$, any its
representations in some basis is completely defined by the actions
of these generators on the same basis. In the case of the typical
representations the matrix elements of the generators in the
reduced basis (4.33) can be obtained by keeping the conditions
(4.1) and (4.41) valid and using the relations (3.1-5). For the
even generators we readily have
\begin{eqnarray*}
{}~~~~~~~~~~~~~~E_{11}(m)_{k}& = &(l_{11}+1)(m)_{k},\\
E_{22}(m)_{k}& = &(l_{12}+l_{22}-l_{11}+2)(m)_{k},\\
E_{12}(m)_{k}& =
&\left([l_{12}-l_{11}][l_{11}-l_{22}]\right)^{1/2}(m)_{k}^{+11},\\
E_{21}(m)_{k}& =
&\left([l_{12}-l_{11}+1][l_{11}-l_{22}-1]\right)^{1/2}(m)_{k}^{-11},\\
E_{33}(m)_{k}& = & (l_{31}+1)(m)_{k},\\ E_{43}(m)_{k}& =
&(l_{32}+l_{42}-l_{31}+2)(m)_{k},\\ E_{34}(m)_{k}& =
&\left([l_{32}-l_{31}][l_{31}-l_{42}]\right)^{1/2}(m)_{k}^{+31},\\
E_{43}(m)_{k}& =
&\left([l_{32}-l_{31}+1][l_{31}-l_{42}-1]\right)^{1/2}(m)_{k}^{-31}.
{}~~~~~~~~~~~~~~~~~~(4.43)
\end{eqnarray*}
\vspace*{2mm}

  As the computations on finding the matrix elements of  $E_{23}$
and $E_{32}$ are too cumbersome, we shall write down here only
the final results. If we assume the formal notation
$$\left|{[x_{1}]...[x_{i}]\over [y_{1}]...[y_{j}]}\right|:=
{[|x_{1}|]...[|x_{i}|]\over [|y_{1}|]...[|y_{j}|]}\eqno(4.44)$$
\\
the generator $E_{23}$ acts on the basis vectors $(m)_{(ab)}$, i.e. on
$(m)_{k}$, as follows
\begin{eqnarray*}
E_{23}(\tilde m)_{(00)} &=& 0,\\[2mm]
E_{23}(\tilde m)_{(10)} &=& -q^{-2}[l_{3-r,3}+l_{s+2,3}+3]
\left|{[l_{3-r,3}-l_{11}][l_{5-s,3}-l_{31}+1]\over
[l_{13}-l_{23}][l_{33}-l_{43}]}\right|^{1/2}
(\tilde m)_{(00)}^{+i2-j2-31},\\[4mm]
E_{23}(\tilde m)_{(ab)} &=&
q^{-2}\sum_{i=max(1,b-r+2)}^{min(2,b-r+3)}
{}~~\sum_{j=max(3,b+s+1)}^{min(4,b+s+2)}(-1)^{(b-1)i+b(j+1)}\\[2mm]
& &\times [l_{i3}+l_{j3}+\left<s\right>-\left<r\right>+3]\\[2mm]
& & \times \left|{[l_{i2}-l_{11}+1][l_{7-j,2}-l_{31}+1]\over
[l_{12}-l_{22}][l_{32}-l_{42}]}\right|^{1/2}
\left|{[l_{i2}-l_{3-i,2}+\left<r\right>-1]\over
[2-\left<r\right>][l_{i3}-l_{3-i,3}+(-1)^{r}]}\right|^{b/2}\\[2mm]
& &\times \left|{[l_{j2}-l_{7-j,2}-\left<s\right>+1]\over
[2-\left<s\right>][l_{j3}-l_{7-j,3}-(-1)^{s}]}\right|^{(1-b)/2}
(\tilde m)_{(10)}^{+i2-j2-31},~~~a+b=2,\\[4mm]
E_{23}(\tilde m)_{(21)} &=&
-q^{-2}\sum_{i=1}^{2}\sum_{j=3}^{4}~~
\sum_{\left<s+j\right>\leq k=0,1\leq \left<r+i\right>}
(-1)^{(1-k)i+kj}\\[2mm]
& &\times
[l_{i3}+l_{j3}-(-1)^{k}\left<i+j+s+r\right>+3]\\[2mm]
& &\times \left|{[l_{i2}-l_{11}+1][l_{7-j,2}-l_{31}+1]\over
[l_{12}-l_{22}][l_{32}-l_{42}]}\right|^{1/2}
\left|{[l_{r2}-l_{i2}+2k-2]\over
[2-k][l_{13}-l_{23}]}\right|^{\left<i+r\right>/2}\\[2mm]
& &\times \left|{[l_{5-s,2}-l_{j2}+2k]\over
[1+k][l_{33}-l_{43}]}\right|^{\left<s+j+1\right>/2}
(\tilde m)_{(1+k,1-k)}^{+i2-j2-31},\\[4mm]
E_{23}(\tilde m)_{(22)} &=&
q^{-2}\sum_{i=1}^{2}\sum_{j=3}^{4}
(-1)^{i+j}
[l_{i3}+l_{j3}+3]\\[2mm]
& &\times \left|{[l_{i3}-l_{11}-1][l_{7-j,3}-l_{31}+3]\over
[l_{13}-l_{23}][l_{33}-l_{43}]}\right|^{1/2}(\tilde m)^{+i2-j2-31}_{(21)},
{}~~~~~~~~~~~~~~~~(4.45)
\end{eqnarray*}
\\[2mm]
while the generator $E_{32}$ has the following matrix elements
\begin{eqnarray*}
E_{32}(\tilde m)_{(00)} &=& -q^{2}\sum_{i=1}^{2}\sum_{j=3}^{4}
  \left|{[l_{i3}-l_{11}][l_{7-j,3}-l_{31}]\over
  [l_{13}-l_{23}][l_{33}-l_{43}]}\right|^{1/2}
  (\tilde m)_{(10)}^{-i2+j2+31},\\[4mm]
  E_{32}(\tilde m)_{(10)} &=& q^{2}\sum_{i=1}^{2}\sum_{j=3}^{4}
{}~~\sum_{\left<r+i\right>\leq k=0,1\leq
\left<s+j\right>}(-1)^{(1-k)i+k(j+1)}\\[2mm]
& &\times \left|{[l_{i2}-l_{11}][l_{7-j,2}-l_{31}]\over
  [l_{12}-l_{22}][l_{32}-l_{42}]}\right|^{1/2}
  \left|{[l_{3-i,3}-l_{i3}+2k-1]\over
[1+k][l_{13}-l_{23}]}\right|^{\left<i+r+1\right>/2}\\[2mm]
& &\times \left|{[l_{7-j,3}-l_{j3}+2k-1]\over
[2-k][l_{33}-l_{43}]}\right|^{\left<s+j\right>/2}
  (\tilde m)_{(2-k,k)}^{-i2+j2+31},\\[4mm]
 E_{32}(\tilde m)_{(ab)}
  &=& -q^{2}\sum_{i=max(1,r-b)}^{min(2,r-b+1)}
 ~~\sum_{j=max(3,5-b-s)}^{min(4,6-b-s)}(-1)^{bi+(1-b)j}\\[2mm]
& &\times \left|{[l_{i2}-l_{11}][l_{7-j,2}-l_{31}]\over
  [l_{12}-l_{22}][l_{32}-l_{42}]}\right|^{1/2}
 \left|{[l_{i2}-l_{3-i,2}-\left<r\right>+1]\over
  [2-\left<r\right>][l_{i3}-l_{3-i,3}-(-1)^{r}]}\right|^{b/2}\\[2mm]
  & &\times \left|{[l_{j2}-l_{7-j,2}+\left<s\right>-1]\over
  [2-\left<s\right>][l_{j3}-l_{7-j,3}+(-1)^{s}]}\right|^{(1-b)/2}
  (\tilde m)_{(21)}^{-i2+j2+31},~~~a+b=2,\\[4mm]
  E_{32}(\tilde m)_{(21)} &=&
-q^{2}(-1)^{r+s}\left|{[l_{r3}-l_{11}-1][l_{s+2,3}-l_{31}+2]\over
[l_{13}-l_{23}][l_{33}-l_{43}]}\right|^{1/2}
(\tilde m)^{-r2+j2+31}_{(22)},~~j=5-s,\\[4mm]
 E_{32}(\tilde m)_{(22)} &=&0.
 ~~~~~~~~~~~~~~~~~~~~~~~~~~~~~~~~~~~~~~~~~~~~~~~~~~~~~~~~
{}~~~~~~~~~~~~~~~~~~~~~~(4.46)
\end{eqnarray*}
where $(\tilde m)_{(ab)}$ are obtained from $(m)_{(ab)}$
by rescaling $$(\tilde m)_{k} = {1\over c_{k}}(m)_{k},
{}~~~~k=0,1,...,15\eqno(4.47)$$.
\vspace*{5mm}
{\large {\bf 5. Conclusion}}\\

    In this work we have constructed all typical representations
of the quantum superalgebras $U_{q}[gl(2/2)]$ at the generic $q$
leaving the coefficients $c_{k}$ (see (4.30-31)) as free
parameters.  The latter can be fixed by some additional
conditions , for example the hermiticity condition. The quantum
typical $U_{q}[gl(2/2)]$-module $W^{q}([m])$ obtained has the
same structure as the classical $gl(2/2)$-module (the module $W$
in Ref. $^{20}$) and can be decomposed into sixteen
$U_{q}[gl(2/2)_{0}]$-fidirmods $V^{q}_{k}([m]_{k})$,
$k=0,1,...,15$.
In general, the method used here is similar to that one of Ref.
$^{20}$. It is not difficult for us to see that for $$c_{k}=1,~~
k=0,1,..,15,$$ we obtain  at $q=1$ the classical results given
in Refs. $^{20}$ (see also Ref. $^{21}$). However, unlike the
latter our approach in this paper avoids the use of the
Clebsch-Gordan
coefficients $^{33}$ which are not always known for higher rank
(quantum and classical) algebras. We hope that the present
approach can be applied for larger quantum superalgebras.
Following the classical programme $^{20,21}$ we can construct atypical
representations of $U_{q}[gl(2/2)]$ at generic $q$ $^{22}$.
Moreover, an extension of the present investigations on the case
when  $q$ is a root of unity is also possible $^{23}$.\\[5mm]
{\bf Acknowledgements}\\

    I am grateful to Prof. Tch. Palev for numerous discussions.
It is a pleasure for me to thank Prof. E. Celeghini and Dr. M.
Tarlini for the kind hospitality at the Florence University where
the present investigations were reported. I am thankful to Profs.
A. Barut, R. Floreanini and V. Rittenberg for useful
discussions.\\

    I would like to thank Prof. Abdus Salam, the International
Atomic Energy Agency and UNESCO for the kind hospitality at the
High Energy- and the Mathematical Section of the International Centre
for Theoretical Physics, Trieste, Italy.\\

    The present work also was partially supported by the National Scientific
Foundation of the Bulgarian Ministry of Science and Higher Education under
the contract F-33.
\newpage
\begin{flushleft}
{\large {\bf 6. References}}
\end{flushleft}
\vspace*{2mm}
\begin{enumerate}
\item L. D. Faddeev, N. Yu. Reshetikhin and L. A. Takhtajan, {\it Algebra
and Analys}, {\bf 1}, 178  (1987).
\item V. D. Drinfel'd, {\it Quantum groups}, in {\it Proceedings of the
International Congress of  Mathematicians}, 1986, Berkeley, vol.
{\bf 1}, 798-820 (The American Mathematical Society, 1987).
\item Yu. I. Manin, {\it Quantum groups and non-commutative geometry}, Centre
des Recherchers Math\'ematiques, Montr\'eal (1988); {\it Topics
in non-commutative geometry}, Princeton University Press,
Princeton, New Jersey (1991).
\item M. Jimbo, {\it Lett. Math. Phys.} {\bf 10}, 63 (1985), {\it ibit} {\bf
11}, 247 (1986).
\item S. I. Woronowicz, {\it Comm. Math. Phys.}, {\bf 111}, 613 (1987).
\item E. K. Sklyanin, {\it Funct. Anal. Appl.}, {\bf 16}, 263 (1982).
\item P. P. Kulish and N. Yu. Reshetikhin, {\it Zapiski nauch. semin.
LOMI} {\bf 101}, 112(1980),(in Russian); English translation: {\it J. Soviet
Math.} {\bf 23}, 2436 (1983).
\item L. C. Biedenharn, {\it J. Phys. A} {\bf 22}, L873 (1989);
A. J. Macfarlane, {\it J. Phys. A} {\bf 22}, 4581 (1989).
\item H. D. Doebner and J. D. Hennig eds., {\it Quantum groups,
Lecture Notes in Physics} {\bf 370} (Springer - Verlag, Berlin
1990).
\item P. P. Kulish ed., {\it Quantum groups,
Lecture Notes in Mathematics} {\bf 1510} (Springer - Verlag, Berlin
1992)
\item E. Celeghini and M. Tarlini eds., {\it Italian Workshop on
quantum groups}, Florence,
February 3-6, 1993, hep-th/9304160; E. Celeghini, {\it Quantum
algebras and Lie groups}, contribution to the Symposium
"{\it Symmetries in Science VI}" in honour of L. C.  Biedenharn,
Bregenz, Austria, August 2-7, 1992 - B. Grubered ed., Plenum, New
York 1992.
\item C. N. Yang  and M. L. Ge eds., {\it Braid groups, knot theory
and statistical mechanics}, World Scientific, Singapore 1989.
\item P. P. Kulish and N. Yu. Reshetikhin, {\it Lett. Math. Phys.} {\bf 18},
143 (1989) ;  E. Celeghini, Tch. Palev and M.  Tarlini, {\it Mod.
Phys. Lett.} {\bf B 5}, 187 (1991).
\item Tch. D. Palev and V. N. Tolstoy, {\it Comm. Math. Phys.}
{\bf 141}, 549 (1991).
\item Yu. I. Manin, {\it Comm. Math. Phys.} {\bf 123}, 163 (1989).
\item M. Chaichian and P. Kulish, {\it Phys. Lett.} {\bf B 234}, 72 (1990).
\item R. Floreanini, V. Spiridonov and L. Vinet, {\it Comm. Math. Phys.}
{\bf 137}, 149 (1991); E. D'Hoker,  R. Floreanini and L. Vinet,
{\it J. Math.  Phys.}, {\bf 32}, 1427 (1991).
\item R. B. Zhang, {\it J. Math.  Phys.}, {\bf 34}, 1236 (1993).
\item V. Kac, {\it Comm. Math. Phys.}, {\bf 53}, 31 (1977); {\it Adv.
Math.} {\bf 26}, 8 (1977); {\it Lecture Notes in Mathematics}
{\bf 676}, 597 (Springer - Verlag, Berlin 1978).
\item A. H. Kamupingene, Nguyen Anh Ky and Tch. D. Palev, {\it J. Math.
Phys.} {\bf 30}, 553 (1989).
\item Tch. Palev and N. Stoilova,
{\it J. Math.  Phys.}, {\bf 31}, 953 (1990).
\item Nguyen Anh Ky, {\it Finite-dimensional representations of
the quantum superalgebra
$U_{q}[gl(2/2)]$: II. Atypical representations at generic $q$},
in preparation.
\item Nguyen Anh Ky, {\it Finite-dimensional representations of
the quantum superalgebra $U_{q}[gl(2/2)]$ at $q$ being roots of
unity}, in preparation.
\item R. Floreanini, D. Leites and L. Vinet, {\it Lett. Math. Phys.}
{\bf 23}, 127 (1991).
\item M. Scheunert, {\it Lett. Math. Phys.} {\bf 24}, 173 (1992);.
\item S. M. Khoroshkin and V. N. Tolstoy, {\it Comm. Math. Phys.}
{\bf 141}, 599 (1991).
\item M. Rosso, {\it Comm. Math. Phys.} {\bf 124}, 307 (1989).
\item M. Rosso, {\it Comm. Math. Phys.} {\bf 117}, 581 (1987).
\item G. Lusztig,
{\it Adv. in Math} {\bf 70}, 237 (1988).
\item I. M. Gel'fand and M. L. Zetlin, {\it Dokl. Akad. Nauk USSR}, {\bf 71},
                   825 (1950), (in Russian); for a detailed
description of the Gel'fand-Zetlin basis see also G. E. Baird and
L. C. Biedenharn, {\it J. Math.  Phys.}, {\bf 4}, 1449 (1963); A.
O. Barut and R. Raczka, {\it Theory of Group Representations and
Applications}, Polish Scientific Publishers, Warszawa, 1980.
\item M. Jimbo, {\it Lecture Notes in Physics}
{\bf 246}, 335 (Springer-Verlag, Berlin 1985); I.V. Cherednik,
{\it Duke Math. Jour.} {\bf 5}, 563 (1987); K. Ueno, T.
Takebayashi and Y. Shibukawa, {\it Lett. Math. Phys.} {\bf 18},
215 (1989).
\newpage
\item V. N. Tolstoy, in ref. $^{9}$ : {\it Quantum Groups,
Lecture Notes in Physics}
{\bf 370} (Springer - Verlag, Berlin 1990), p. 118.
\item V. A. Groza, I. I. Kachurik and A. U. Klimyk,
{\it J. Math.  Phys.}, {\bf 31}, 2769 (1990).
\end{enumerate}
\newpage
\begin{flushleft}
{\bf Appendix}\\
\end{flushleft}

  The induced basis (4.20) is expressed in terms of the reduced
basis  through the following invert relation
\begin{eqnarray*}
\left|1,0,0,0;(m)\right>& = &\left\{{1\over c_{1}}q^{2(l+p+1)}
\left({[l_{13}-l_{11}+1][l_{31}-l_{43}]\over [2l+1][2l'+1]}
\right)^{1/2}(m)_{1}^{-11+31}\right.\\[2mm] &   &-{1\over
c_{2}}q^{2(-l+p)} \left({[l_{11}-l_{23}-1][l_{31}-l_{43}]\over
[2l+1][2l'+1]} \right)^{1/2}(m)_{2}^{-11+31}\\[2mm] &   &+{1\over
c_{3}}q^{2(l+p+1)} \left({[l_{13}-l_{11}+1][l_{33}-l_{31}]\over
[2l+1][2l'+1]} \right)^{1/2}(m)_{3}^{-11+31}\\[2mm] &   &\left.
-{1\over c_{4}}q^{2(-l+p)}
\left({[l_{11}-l_{23}-1][l_{33}-l_{31}]\over [2l+1][2l'+1]}
\right)^{1/2}(m)_{4}^{-11+31}\right\},\\[2mm]
\left|0,1,0,0;(m)\right>& = &\left\{-{1\over c_{1}}q^{2}
\left({[l_{11}-l_{23}][l_{31}-l_{43}]\over [2l+1][2l'+1]}
\right)^{1/2}(m)_{1}^{+31}\right.\\[2mm] &   &-{1\over c_{2}}q^{2}
\left({[l_{13}-l_{11}][l_{31}-l_{43}]\over [2l+1][2l'+1]}
\right)^{1/2}(m)_{2}^{+31}\\[2mm] &   &-{1\over c_{3}}q^{2}
\left({[l_{11}-l_{23}][l_{33}-l_{31}]\over [2l+1][2l'+1]}
\right)^{1/2}(m)_{3}^{+31}\\[2mm] &   &\left. -{1\over c_{4}}q^{2}
\left({[l_{13}-l_{11}][l_{33}-l_{31}]\over [2l+1][2l'+1]}
\right)^{1/2}(m)_{4}^{+31}\right\},\\[2mm]
\left|0,0,1,0;(m)\right>& = &\left\{{1\over c_{1}}q^{2(l+p-l'-p')}
\left({[l_{13}-l_{11}+1][l_{33}-l_{31}+1]\over [2l+1][2l'+1]}
\right)^{1/2}(m)_{1}^{-11}\right.\\[2mm] &   &-{1\over
c_{2}}q^{2(-l+p-l'-p'-1)}
\left({[l_{11}-l_{23}-1][l_{33}-l_{31}+1]\over [2l+1][2l'+1]}
\right)^{1/2}(m)_{2}^{-11}\\[2mm] &   &-{1\over
c_{3}}q^{2(l+p+l'-p'+1)}
\left({[l_{13}-l_{11}+1][l_{31}-l_{43}-1]\over [2l+1][2l'+1]}
\right)^{1/2}(m)_{3}^{-11}\\[2mm] &   &\left. +{1\over
c_{4}}q^{2(-l+p+l'-p')}
\left({[l_{11}-l_{23}-1][l_{31}-l_{43}-1]\over [2l+1][2l'+1]}
\right)^{1/2}(m)_{4}^{-11}\right\},\\[2mm]
\left|0,0,0,1;(m)\right>& = &\left\{-{1\over c_{1}}q^{-2(l'+p')}
\left({[l_{11}-l_{23}][l_{33}-l_{31}+1]\over [2l+1][2l'+1]}
\right)^{1/2}(m)_{1}\right.\\[2mm] &   &-{1\over
c_{2}}q^{-2(l'+p')}
\left({[l_{13}-l_{11}][l_{33}-l_{31}+1]\over [2l+1][2l'+1]}
\right)^{1/2}(m)_{2}\\[2mm] &   &+{1\over c_{3}}q^{2(l'-p'+1)}
\left({[l_{11}-l_{23}][l_{31}-l_{43}-1]\over [2l+1][2l'+1]}
\right)^{1/2}(m)_{3}\\[2mm] &   &\left. +{1\over
c_{4}}q^{2(l'-p'+1)} \left({[l_{13}-l_{11}][l_{31}-l_{43}-1]\over
[2l+1][2l'+1]} \right)^{1/2}(m)_{4}\right\},\\[2mm]
\left|1,0,1,0;(m)\right>& = &\left\{{1\over c_{5}}q^{2(2l+2p+1)}
\left({[l_{13}-l_{11}+1][l_{13}-l_{11}+2]\over [2l+1][2l+2]}
\right)^{1/2}(m)_{5}^{-11-11+31}\right.\\[2mm] &   &-{1\over
c_{6}}q^{2(2p-1)}
\left({[2][l_{13}-l_{11}+1][l_{11}-l_{23}-1]\over [2l][2l+2]}
\right)^{1/2}(m)_{6}^{-11-11+31}\\[2mm] &   &\left. +{1\over
c_{7}}q^{2(-2l+2p-1)}
\left({[l_{11}-l_{23}-2][l_{11}-l_{23}-1]\over [2l][2l+1]}
\right)^{1/2}(m)_{7}^{-11-11+31}\right\},\\[2mm]
\left|1,0,0,1;(m)\right>& = &\left\{-{1\over c_{5}}q^{2(l+p+2)}
\left({[l_{13}-l_{11}+1][l_{11}-l_{23}]\over [2l+1][2l+2]}
\right)^{1/2}(m)_{5}^{-11+31}\right.\\[2mm] &   &-{1\over
c_{6}}{[2]^{1/2}q^{2(p+1)}(q^{2l+2}[l_{13}-l_{11}]-
q^{-2l-2}[l_{11}-l_{23}-1])\over (q^{2}+q^{-2})
\left([2l][2l+2]\right)^{1/2}}(m)_{6}^{-11+31}\\[2mm] &
&+{1\over c_{7}}q^{2(-l+p+1)}
\left({[l_{13}-l_{11}][l_{11}-l_{23}-1]\over [2l][2l+1]}
\right)^{1/2}(m)_{7}^{-11+31}\\[2mm] &   &+{1\over
c_{8}}q^{-2(l'+p')}
\left({[l_{33}-l_{31}+1][l_{31}-l_{43}]\over [2l'+1][2l'+2]}
\right)^{1/2}(m)_{8}^{-11+31}\\[2mm] &   &-{1\over
c_{9}}{[2]^{1/2}q^{2(-p'+1)}(q^{2l'+2}[l_{31}-l_{43}-1]-
q^{-2l'-2}[l_{33}-l_{31}])\over
(q^{2}+q^{-2})\left([2l'][2l'+2]\right)^{1/2}}(m)_{9}^{-11+31}\\[2mm]
&   &\left. -{1\over c_{10}}q^{2(l'-p'+1)}
\left({[l_{33}-l_{31}][l_{31}-l_{43}-1]\over [2l'][2l'+1]}
\right)^{1/2}(m)_{10}^{-11+31}\right\},\\[2mm]
\left|0,1,1,0;(m)\right>& = &\left\{-{1\over c_{5}}q^{2(l+p+1)}
\left({[l_{13}-l_{11}+1][l_{11}-l_{23}]\over [2l+1][2l+2]}
\right)^{1/2}(m)_{5}^{-11+31}\right.\\[2mm] &   &-{1\over
c_{6}}{[2]^{1/2}q^{2p}(q^{2l+2}[l_{13}-l_{11}]-
q^{-2l-2}[l_{11}-l_{23}-1])\over
(q^{2}+q^{-2})\left([2l][2l+2]\right)^{1/2}}(m)_{6}^{-11+31}\\[2mm]
&   &+{1\over c_{7}}q^{2(-l+p)}
\left({[l_{13}-l_{11}][l_{11}-l_{23}-1]\over [2l][2l+1]}
\right)^{1/2}(m)_{7}^{-11+31}\\[2mm] &   &-{1\over
c_{8}}q^{2(-l'-p'+1)} \left({[l_{33}-l_{31}+1][l_{31}-l_{43}]\over
[2l'+1][2l'+2]} \right)^{1/2}(m)_{8}^{-11+31}\\[2mm] &
&+{1\over
c_{9}}{[2]^{1/2}q^{2(-p'+2)}(q^{2l'+2}[l_{31}-l_{43}-1]-
q^{-2l'-2}[l_{33}-l_{31}])\over
(q^{2}+q^{-2})\left([2l'][2l'+2]\right)^{1/2}}(m)_{9}^{-11+31}\\[2mm]
&   &\left. +{1\over c_{10}}q^{2(l'-p'+2)}
\left({[l_{33}-l_{31}][l_{31}-l_{43}-1]\over [2l'][2l'+1]}
\right)^{1/2}(m)_{10}^{-11+31}\right\},\\[2mm]
\left|0,1,0,1;(m)\right>& = &\left\{{1\over c_{5}}q^{2}
\left({[l_{11}-l_{23}][l_{11}-l_{23}+1]\over [2l+1][2l+2]}
\right)^{1/2}(m)_{5}^{+31}\right.\\[2mm] &   &+{1\over c_{6}}q^{2}
\left({[2][l_{11}-l_{23}][l_{13}-l_{11}]\over [2l][2l+2]}
\right)^{1/2}(m)_{6}^{+31}\\[2mm] &   &\left. +{1\over c_{7}}q^{2}
\left({[l_{13}-l_{11}-1][l_{13}-l_{11}]\over [2l][2l+1]}
\right)^{1/2}(m)_{7}^{+31}\right\},\\[2mm]
\left|1,1,0,0;(m)\right>& = &\left\{{1\over c_{8}}q^{2}
\left({[l_{31}-l_{43}][l_{31}-l_{43}+1]\over [2l'+1][2l'+2]}
\right)^{1/2}(m)_{8}^{-11+31+31}\right.\\[2mm] &   &+{1\over
c_{9}}q^{2} \left({[2][l_{31}-l_{43}][l_{33}-l_{31}]\over
[2l'][2l'+2]} \right)^{1/2}(m)_{9}^{-11+31+31}\\[2mm] &   &\left.
+{1\over c_{10}}q^{2} \left({[l_{33}-l_{31}-1][l_{33}-l_{31}]\over
[2l'][2l'+1]} \right)^{1/2}(m)_{10}^{-11+31+31}\right\},\\[2mm]
\left|0,0,1,1;(m)\right>& = &\left\{{1\over c_{8}}q^{-2(2l'+2p'+1)}
\left({[l_{33}-l_{31}+1][l_{33}-l_{31}+2]\over [2l'+1][2l'+2]}
\right)^{1/2}(m)_{8}^{-11}\right.\\[2mm] &   &-{1\over
c_{9}}q^{2(-2p'+1)} \left({[2][l_{33}-l_{31}+1][l_{31}-l_{43}-1]\over
[2l'][2l'+2]} \right)^{1/2}(m)_{9}^{-11}\\[2mm] &   &\left.
+{1\over c_{10}}q^{2(2l'-2p'+1)}
\left({[l_{31}-l_{43}-2][l_{31}-l_{43}-1]\over [2l'][2l'+1]}
\right)^{1/2}(m)_{10}^{-11}\right\},\\[2mm]
\left|1,1,1,0;(m)\right>& = &\left\{-{1\over c_{11}}q^{2(l+p+1)}
\left({[l_{13}-l_{11}+1][l_{31}-l_{43}]\over [2l+1][2l'+1]}
\right)^{1/2}(m)_{1}^{-11-11+31+31}\right.\\[2mm] &   &+{1\over
c_{12}}q^{2(-l+p)} \left({[l_{11}-l_{23}-1][l_{31}-l_{43}]\over
[2l+1][2l'+1]} \right)^{1/2}(m)_{2}^{-11-11+31+31}\\[2mm] &
&-{1\over c_{13}}q^{2(l+p+1)}
\left({[l_{13}-l_{11}+1][l_{33}-l_{31}]\over [2l+1][2l'+1]}
\right)^{1/2}(m)_{3}^{-11-11+31+31}\\[2mm] &   &\left. +{1\over
c_{14}}q^{2(-l+p)} \left({[l_{11}-l_{23}-1][l_{33}-l_{31}]\over
[2l+1][2l'+1]}
\right)^{1/2}(m)_{14}^{-11-11+31+31}\right\},\\[2mm]
\left|1,1,0,1;(m)\right>& = &\left\{{1\over c_{11}}q^{2}
\left({[l_{11}-l_{23}][l_{31}-l_{43}]\over [2l+1][2l'+1]}
\right)^{1/2}(m)_{11}^{-11+31+31}\right.\\[2mm] &   &+{1\over
c_{12}}q^{2} \left({[l_{13}-l_{11}][l_{31}-l_{43}]\over [2l+1][2l'+1]}
\right)^{1/2}(m)_{12}^{-11+31+31}\\[2mm] &   &+{1\over c_{13}}q^{2}
\left({[l_{11}-l_{23}][l_{33}-l_{31}]\over [2l+1][2l'+1]}
\right)^{1/2}(m)_{3}^{-11+31+31}\\[2mm] &   &\left. +{1\over
c_{14}}q^{2} \left({[l_{13}-l_{11}][l_{33}-l_{31}]\over [2l+1][2l'+1]}
\right)^{1/2}(m)_{14}^{-11+31+31}\right\},\\[2mm]
\left|1,0,1,1;(m)\right>& = &\left\{{1\over c_{11}}q^{2(l+p-l'-p')}
\left({[l_{13}-l_{11}+1][l_{33}-l_{31}+1]\over [2l+1][2l'+1]}
\right)^{1/2}(m)_{11}^{-11-11+31}\right.\\[2mm] &   &-{1\over
c_{12}}q^{2(-l+p-l'-p'-1)}
\left({[l_{11}-l_{23}-1][l_{33}-l_{31}+1]\over [2l+1][2l'+1]}
\right)^{1/2}(m)_{12}^{-11-11+31}\\[2mm] &   &-{1\over
c_{13}}q^{2(l+p+l'-p'+1)}
\left({[l_{13}-l_{11}+1][l_{31}-l_{43}-1]\over [2l+1][2l'+1]}
\right)^{1/2}(m)_{13}^{-11-11+31}\\[2mm] &   &\left. +{1\over
c_{14}}q^{2(-l+p+l'-p')}
\left({[l_{11}-l_{23}-1][l_{31}-l_{43}-1]\over [2l+1][2l'+1]}
\right)^{1/2}(m)_{14}^{-11-11+31}\right\},\\[2mm]
\left|0,1,1,1;(m)\right>& = &\left\{-{1\over c_{11}}q^{-2(l'+p')}
\left({[l_{11}-l_{23}][l_{33}-l_{31}+1]\over [2l+1][2l'+1]}
\right)^{1/2}(m)_{11}^{-11+31}\right.\\[2mm] &   &-{1\over
c_{12}}q^{-2(l'+p')}
\left({[l_{13}-l_{11}][l_{33}-l_{31}+1]\over [2l+1][2l'+1]}
\right)^{1/2}(m)_{12}^{-11+31}\\[2mm] &   &+{1\over
c_{13}}q^{2(l'-p'+1)} \left({[l_{11}-l_{23}][l_{31}-l_{43}-1]\over
[2l+1][2l'+1]} \right)^{1/2}(m)_{13}^{-11+31}\\[2mm] &   &\left.
+{1\over c_{14}}q^{2(l'-p'+1)}
\left({[l_{13}-l_{11}][l_{31}-l_{43}-1]\over [2l+1][2l'+1]}
\right)^{1/2}(m)_{14}^{-11+31}\right\},\\[2mm]
\left|1,1,1,1;(m)\right>& = &{1\over c_{15}}(m)^{-11-11+31+31}.
\end{eqnarray*}
\end{document}